\documentclass[aps,prx,twocolumn,superscriptaddress,nofootinbib,floatfix,letterpaper]{revtex4-1}

\usepackage{draft,url} 
\usepackage{hyperref,upgreek,amssymb}
\usepackage{graphicx,color}
\usepackage{mathtools}
\usepackage{float}
\usepackage{mciteplus}
\usepackage{blindtext}
\DeclareFontFamily{OT1}{pzc}{}
\DeclareFontShape{OT1}{pzc}{m}{it}{<-> s * [1.10] pzcmi7t}{}
\DeclareMathAlphabet{\mathpzc}{OT1}{pzc}{m}{it}

\numberwithin{equation}{section}
\newcommand{\ii}{\hspace{1pt}\mathrm{i}\hspace{1pt}}

\def\be#1\ee{\begin{align}#1\end{align}}
\newcommand\nn{\nonumber}

\newcommand{\ZZ}{\mathbb{Z}}
\newcommand{\calF}{\mathcal{F}}
\newcommand{\calO}{\mathcal{O}}

\newcommand{\fh}{\frac{1}{2}}

\begin{document}

\begin{titlepage}

\title{Topological Transition on the Conformal Manifold}

\author{Wenjie Ji$^{a}$,  Shu-Heng Shao$^b$, and Xiao-Gang Wen$^a$
}

\address{${}^a$Department of Physics, Massachusetts Institute of Technology,\\ Cambridge, Massachusetts 02139, USA}
\address{${}^b$School of Natural Sciences, Institute for Advanced Study,\\
Princeton, New Jersey 08540, USA}

\begin{abstract}
Despite great successes in the study of gapped phases, a
comprehensive understanding of the gapless phases and their transitions is
still under developments.  In this paper, we study a general phenomenon  in the
space of (1+1)$d$ critical phases with fermionic degrees of freedom described
by a continuous family of conformal field theories (CFT), a.k.a.\ the conformal
manifold.  Along a one-dimensional locus on the conformal manifold, there can
be a transition point, across which the fermionic  CFTs on the two sides differ
by stacking an invertible fermionic topological order (IFTO), point-by-point
along the locus.  At every point on the conformal manifold, the order and
disorder operators have power-law two-point functions, but their critical
exponents cross over with each other at the transition point, where stacking
the IFTO leaves the fermionic CFT unchanged.  We call this continuous transition
on the fermionic conformal manifold a {\it topological transition}.  By gauging
the fermion parity, the IFTO stacking becomes a Kramers-Wannier duality between
the corresponding bosonic CFTs.   Both the IFTO stacking and the
Kramers-Wannier duality are induced by the electromagnetic duality of the
(2+1)$d$ $\mathbb{Z}_2$ topological order.  We provide several examples of
topological transitions, including  the familiar Luttinger model of spinless
fermions (i.e.\ the $c=1$ massless Dirac fermion with the Thirring
interaction), and a new class of $c=2$ examples describing $U(1)\times SU(2)$-protected gapless phases.
\end{abstract}

\pacs{}

\maketitle

\end{titlepage}

\section{Introduction}

The study of gapped topological ordered phases has been systematically
developed over the last thirty years. The field has moved to the direction of
studying (1) the critical theories between topological phases, or between a
topological phase and a trivial phase. (2) the gapless phases without
quasiparticles.  In recent years, there are several progresses in studying
transitions between topological phases. One is to consider the continuous
transition between gapped trivial phase and gapped topological ordered phases
that do not involve any spontaneously symmetry
breaking\cite{WW9301,CFW9349,W0050,W0213,SBS0407} (but may be viewed as the
spontaneously breaking of emergent higher-symmetry\cite{W181202517} in the sense of \cite{Kapustin:2014gua,Gaiotto:2014}). The
critical points between symmetry protected topological phases are also examples
of continuous  transitions that do not involve any spontaneously symmetry
breaking\cite{Chen:2013inn,Tsui:2015dja,Tsui:2015dmg,Bultinck:2019zzo,bi2019adventure}.  
With or without the  spontaneously
symmetry breaking, the critical points for continuous phase transition
are often gapless states without well defined
quasiparticles \cite{RW0171,HSF0437,SBS0407}.
 Moreover, some gapless phases can be strongly
correlated and have no well-defined quasiparticles down to  zero energy,
such as the large $N$ QED in (2+1)$d$, certain $U(1)$ spin
liquid\cite{RW0171,HSF0437}, and  QED in (3+1)$d$.

To characterize gapless phases without quasiparticles, inspired by the success in gapped phases, we start with the question whether 
they can be topologically nontrivial.
A  far from exhaustive list of reference are \cite{Ji19TQT,PhysRevB.83.174409,cheng2011majorana,fidkowski2011majorana,sau2011number,kraus2013majorana,keselman2015gapless,iemini2015localized,montorsi2017symmetry,ruhman2017topological,jiang2018symmetry,PhysRevB.98.214501}. To give an example, one construction is
to impose symmetries and to decorate domain walls in gapless phases by symmetry
charges, in analogy of construction of symmetry protected topological phases 
\cite{scaffidi2017gapless,parker2018topological}.  Overall, this line of
questions is hard and a universal understanding is in demand. 

More generally, we would like to understand how to distinguish the topological
nature in a gapless phase. It is proposed recently to use the topological edge mode in gapless phases\cite{verresen2017one,verresen2018topology,Verresen:2019igf}. 
 A numerical success has been made in identifying ground state degeneracy in (1+1)$d$ gapless models with open boundary condition. 
 The energy splitting scales with the system size either exponentially or with power law with a large exponent. 
The low energy theories are conformal field theories (CFT) without symmetry-preserving relevant operator. It is also proposed in the above literature that there can be topological invariant defined in the gapless bulk, based on different ways to assign symmetry charges to the non-local operators in the low energy effective field theory.

In this paper, we introduce a concrete field-theoretic setup where the topological nature of a gapless state changes as we dial the parameters of the model.  
We consider a one-parameter family of CFTs  in (1+1)$d$ with fermionic degrees of freedom, labeled by an exactly marginal coupling $g$.  
The latter parametrizes a one-dimensional locus of the conformal manifold where every point defines a fermionic (spin) CFT with the same central charge. 
We will be interested in the scenario where the CFT ${\cal F}[-g]$ with negative coupling differs from that ${\cal F}[g]$ with positive coupling by an invertible fermionic topological order (IFTO) ({\it a.k.a.} the Arf invariant), point-by-point for all couplings $g$. 
In other words, the gapless state ${\cal F}[g]$ can be viewed as
the other state ${\cal F}[-g]$ stacked with a (1+1)$d$ $p$-wave superconducting state \cite{Kitaev:2001kla}.

The CFT ${\cal F}[0]$ at the transition point $g=0$ enjoys an enhanced $\bZ_2^{\rm IFTO}$ global symmetry: it  is invariant under stacking an IFTO.   
Away from the transition point, the $\bZ_2^{\rm IFTO}$ action is not a symmetry, but maps ${\cal F}[g]$ to ${\cal F}[-g]$ by stacking an IFTO. 
Since the CFTs on the two sides of ${\cal F}[0]$ differ by a (1+1)$d$ topological order, we will call the transition ${\cal F}[-g]\to {\cal F}[0]\to {\cal F}[g]$ a {\it topological transition} on the conformal manifold.  
The simplest example is a $c=1$ massless Dirac fermion  with a quartic fermion interaction ({\it i.e.} the Thirring coupling)   \cite{Karch:2019lnn}, which can be equivalently described by the Luttinger model of spinless fermions.  
We will also discuss a $c=2$ example with  $U(1)\times SU(2)$ global symmetry.

The topological transition is similar to the standard second-order phase
transition, and yet it is different in many aspects.  It is similar in that the
gapless states on both sides differ by a topological order, much as a  (1+1)$d$
massive Majorana fermions  with $m>0$ and $m<0$ do.  However, the gap in a
topological transition is always exactly zero, and the CFT data ({\it e.g.} the spectrum and the quantum numbers of the
local operators) change continuously as we vary the
exactly marginal coupling $g$.  Another resemblance is that while the order and
the disorder operators both have  power-law two-point functions along the
topological transition, their critical exponents cross over with each other at
the transition point ${\cal F}[0]$.  This is to be contrasted with the standard
order-disorder phase transition where in one phase the order operators have
asymptotic constant correlations while the disorder operators have
exponentially decaying correlations, and vice versa in the other phase.

The topological transition ${\cal F}[-g] \to {\cal F}[0]\to {\cal F}[g]$ has a
parallel story in the bosonized picture.  The bosonization and fermionization
maps in (1+1)$d$ quantum field theory have been recently revisited from a more
modern point of view
\cite{Gaiotto:2015zta,Bhardwaj:2016clt,Kapustin:2017jrc,Thorngren:2018bhj,Karch:2019lnn,yujitasi},
which we will review in Section \ref{sec:bosferm}.  Let ${\cal B}[g]$ be the
bosonization of ${\cal F}[g]$ by gauging the fermion parity, then ${\cal B}[g]$
and ${\cal B}[-g]$ differ by a $\bZ_2^{\cal B}$ orbifold
\cite{Kapustin:2014gua,Karch:2019lnn}.  In particular, the bosonic transition
point ${\cal B}[0]$ is self-dual under the $\bZ_2^{\cal B}$ orbifold, which
generalizes the notion of the Kramers-Wannier duality
\cite{PhysRev.60.252,Monastyrsky:1978kp,savit1980duality} to more general bosonic (non-spin) CFT than the
Ising CFT.  By exploiting  our knowledge on the bosonic conformal manifold, we
produce several examples of topological transitions for fermionic CFTs.

The line of CFTs under consideration can also be realized by (1+1)$d$ lattice
models where  $\bZ_2^{\rm IFTO}$ is a symmetry of the models.  But such a
$\bZ_2^{\rm IFTO}$ symmetry has a 't Hooft anomaly and is not on-site in the
lattice models.\footnote{On the lattice model, (In the bosonic model, the $\ZZ_2$ orbifold is realized by translation by half a site, mapping site degrees of freedom to link degrees of freedom. The simplest example is the Ising model.) }  We can also realize the line of CFTs by the boundaries of (2+1)$d$ 
$\bZ_2$ topological order ({\it i.e.} $\bZ_2$ gauge theory), where $\bZ_2^{\rm
IFTO}$ is the on-site symmetry of the 2+1 models that exchange the $\bZ_2$-charge
$e$ and the $\bZ_2$-vortex $m$.  In the above two families of CFTs with $\bZ_2^{\rm IFTO}$
symmetry, ${\cal F}[g]$ and ${\cal F}[-g]$ represent the two degenerate ground
states from spontaneous $\bZ_2^{\rm IFTO}$-symmetry breaking.

\begin{figure*}
\includegraphics[scale=1]{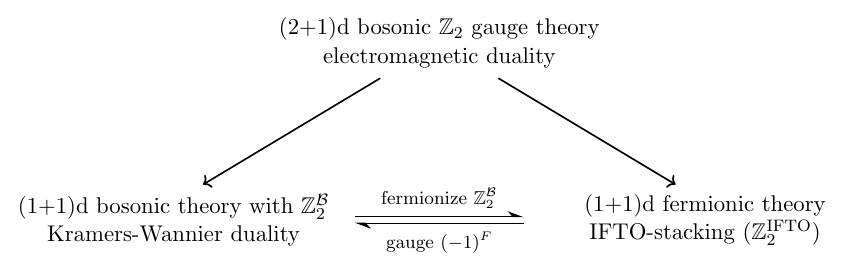}
\caption{We can couple the (2+1)$d$ bosonic $\ZZ_2$ gauge theory to   a (1+1)$d$ bosonic theory with a non-anomalous $\bZ_2^{\cal B}$ symmetry, or  to a (1+1)$d$ fermionic theory. The (2+1)$d$ electromagnetic duality implements either the Kramer-Wannier duality $(\bZ_2^{\cal B}$ orbifold) when the (1+1)$d$ boundary is bosonic, or the IFTO stacking ($\bZ_2^{\rm IFTO})$ when the boundary is fermionic.}\label{fig:triangle}
\end{figure*}

Both the Kramers-Wannier duality of (1+1)$d$ bosonic theories and the IFTO
stacking $\bZ_2^{\rm IFTO}$ action of (1+1)$d$ fermionic theories are
intimately related to electromagnetic duality in the (2+1)$d$ bosonic untwisted
$\bZ_2$ gauge theory \cite{rey1991self}.  Indeed, the electromagnetic duality
of the $\bZ_2$ gauge theory  extends to the boundary with local bosons (with an
anomaly-free $\ZZ_2^{\cal B}$ twisting) as the Kramers-Wannier duality
\cite{Severa:2002qe,Freed:2018cec}.  On the other hand, when the (1+1)$d$
boundary has local fermions, the electromagnetic duality extends to stacking an
IFTO \cite{Ji:2019eqo}.  Purely from the (1+1)$d$ boundary point of view, the
Kramers-Wannier duality is related to the $\bZ_2^{\rm IFTO}$ stacking via
bosonization/fermionization (as we will discuss in Section \ref{sec:bosferm}),
which provides a direct translation between the two extensions.
This is analogous to the relation between the (3+1)$d$ Maxwell theory and the (2+1)$d$ particle-vortex dualities \cite{Witten:2003ya,Seiberg:2016gmd}. 
 See Figure
\ref{fig:triangle}.\footnote{Throughout the paper, we will assume that the
gravitational anomaly of the (1+1)$d$ fermionic theory is $c_L - c_R=  0 $ mod
8, and the $(-1)^F$ fermion parity has no 't Hooft anomaly.}

The generalized Kramers-Wannier duality of a bosonic CFT ${\cal B}_0$ can be implemented by the non-invertible duality defect line in the (1+1)$d$ spacetime \cite{Frohlich:2004ef,Frohlich:2006ch,Frohlich:2009gb,Aasen:2016dop,Bhardwaj:2017xup,Chang:2018iay,Buican:2017rxc}. 
We give a detailed analysis of the duality defect in several examples, and discuss its relation to the $\bZ_2^{\rm IFTO}$ symmetry defect of the corresponding fermionic CFT ${\cal F}_0$.

This paper is organized as follows. 
In Section \ref{sec:Ftransition}, we review the IFTO and discuss general features of topological transitions on  fermionic conformal manifolds.  
We also discuss the interpretation of stacking an IFTO from the (2+1)$d$ $\bZ_2$ gauge theory point of view. 
In Section \ref{sec:bosferm}, we review the bosonization and the  fermionization procedures in (1+1)$d$. 
In particular, we show that two fermionic theories differ by an IFTO if and only if their bosonized theories are related by a $\bZ_2$ orbifold. 
In Section \ref{sec:Btransition}, we discuss several examples of bosonic conformal manifolds where CFTs are related by a $\bZ_2$ orbifold, which includes the $c=1$ compact boson $S^1$, the $c=1$ orbifold theory $S^1/\bZ_2$, as well as a $c=2$ $T^2$ CFT example. 
These bosonic examples pave the way for the topological transition of their fermionizations, including the $c=1$ massless Thirring model ({\it a.k.a.} the Luttinger liquid), which we discuss in Section \ref{sec:Fmodels}. 
We will also discuss a $c=2$ fermionic CFT describing a spin$-\fh$ gapless phase beyond the Luttinger liquid type. 
In Section \ref{app:ext}, we discuss how a symmetry defect in a fermionic theory becomes a duality defect under bosonization. 
We end with several future directions in Section \ref{sec:conclu}.

\section{Topological Transition on the Fermionic Conformal Manifold}\label{sec:Ftransition}

In this paper we consider gapless states described by CFTs with fermionic degrees of freedom.  In particular, we focus on CFTs with exactly marginal deformations.  
By turning on the  exactly marginal deformations, one generates a continuous family of CFTs parametrized by the exactly marginal couplings.  
The space of this family of CFTs is called  a  \textit{conformal manifold}.  
In this section, we describe a general phenomenon where along a one-dimensional slice of the conformal manifold, there is a transition point across which the CFTs on the two sides differ by an IFTO stacking. 

\subsection{Invertible Fermionic Topological Order}

We start by reviewing the IFTO, which is  the Kitaev chain of $p$-wave superconductor \cite{Kitaev:2001kla} (see \cite{Kapustin:2014dxa,Kapustin:2014gua,Shiozaki:2016zjg,Dijkgraaf:2018vnm,Senthil:2018cru,Karch:2019lnn} for further discussions). 

Given a Riemann surface $\Sigma$ with a spin structure $\rho$, the Arf invariant ${\rm Arf}[\rho]$ is defined as \cite{atiyah}
\ie
{\rm Arf}[\rho]=
\begin{cases}
1\,,~~~~~\text{if}~~~\text{$\rho$ is odd}\,,\\
0\,,~~~~~\text{if}~~~\text{$\rho$ is even}\,.
\end{cases}
\fe
On a genus-$g$ Riemann surface, there are $2^{g-1}(2^g-1)$ odd spin structures and $2^{g-1}(2^g+1)$ even spin structures. 

A $\bZ_2$ gauge connection $s\in H^1(\Sigma,\bZ_2)$ is specified by the holonomy $\oint_\gamma s \in \{0,1\}$ around each cycle $\gamma$ of $\Sigma$. 
Given a spin structure $\rho$ and a $\bZ_2$ connection, we can construct a new spin structure modified by the $\bZ_2$ twist. We will denote this new spin structure by $s+\rho$.

Using the Arf invariant, we can define a (1+1)-dimensional IFTO. 
The partition function of this IFTO on  a Riemann surface   with spin structure $\rho$ is simply given by
\ie\label{FSPT}
Z_{\rm IFTO} [\rho]  = e^{\ii \pi {\rm Arf}[\rho] }\,.
\fe
This IFTO is protected by the $(-1)^F$ symmetry.  
The sign  of $Z_{\text{IFTO}}[\rho]$,  compared to the trivially  fermionic gapped phase $Z[\rho]=1$, measures the parity of the number of Majorana zero modes\cite{Kitaev:2001kla}, or the parity change of the number of negative energy eigenstates. 
 If we stack two IFTOs together, then it becomes  a trivial phase. Let the $\bZ_2$ background gauge field of $(-1)^F$ be $S$.  The partition function of the IFTO coupled to a background $(-1)^F$ gauge field $S$ is  $Z_{\rm IFTO}[S+\rho] = e^{\ii \pi {\rm Arf}[S+\rho]}$.

\subsection{Topological Transition of the Fermionic Gapless States}\label{sec:phase}

We start with a general discussion on fermionic states in (1+1)
dimensions described by a fermionic quantum field theory (QFT) $\cal F$.\footnote{A QFT whose partition function requires a choice of the spin structure is called a spin or a fermionic QFT, such as the Majorana fermion.  By contrast, a QFT whose partition function does not require a choice of the spin structure is called a non-spin or a bosonic QFT, such as the Ising CFT.}   
From ${\cal F}$ we can construct another
fermionic theory ${\cal F}'$ by stacking with an IFTO $Z_{\rm IFTO}$:
\ie\label{FFp}
Z_{{\cal F}' }  [\rho ]=  Z_{\rm IFTO}[\rho]Z_{{\cal F}}  [\rho ]\,.
\fe
The partition functions for ${\cal F}$ and ${\cal F}'$ are identical  on
Riemann surfaces with even spin structure, but differ by a sign for odd spin
structure.

On a spacetime torus with complex structure moduli $q=\exp(2\pi \ii \tau)$, a fermion system has four partition functions
\ie\label{ZF1}
&Z_{\cal F} [AE]  = \text{Tr}_{\cal H_A}\left[  \,{1+(-1)^F\over 2}  \,q^{ H+K\over2 } \bar q^{ H-K\over2 }  \,\right]\,,\\
&Z_{\cal F} [AO]  = \text{Tr}_{\cal H_A}\left[ \, {1-(-1)^F\over 2}   \,q^{ H+K\over2 } \bar q^{ H-K\over2 }    \,\right]\,,\\
&Z_{\cal F} [PE]  = \text{Tr}_{\cal H_P}\left[  \,  {1+(-1)^F\over 2}   \,q^{ H+K\over2 } \bar q^{ H-K\over2 }\, \right]\,,\\
&Z_{\cal F} [PO]  = \text{Tr}_{\cal H_P}\left[ \, {1-(-1)^F\over 2}   \,q^{ H+K\over2 } \bar q^{ H-K\over2 }  \,\right ]\,,
\fe
where ${\cal H}_P$ (${\cal H}_A$) is the Hilbert space with periodic (antiperiodic) boundary condition for the fermions.\footnote{In the high energy terminology, ${\cal H}_P$ is the Ramond sector while ${\cal H}_A$ is the Neveu-Schwarz sector.}  
$H$ and $K$ are the eigenvalues of the Hamiltonian and momentum in the corresponding Hilbert space. 
In a CFT with central charges $c=c_L=c_R$, $H$ and $K$ are related to the conformal weights as $h - {c\over24} =  {H+K\over2}$ and $\bar h -{c\over24} = {H-K\over2}$. 
 Here $E$ and $O$ stand for the $(-1)^F$-even and $(-1)^F$-odd sectors, respectively.

Alternatively, we may define the torus partition functions for fermion systems
through the space-time path integral, which also include four types,
$Z_{\cal F}[AA]$, $Z_{\cal F}[AP]$, $Z_{\cal F}[PA]$, and $Z_{\cal F}[PP]$.  Here the
first and second subscription $P$ or $A$ corresponds the periodic or
anti-periodic boundary condition for fermions in $x$ and $t$ direction,
respectively. The two sets of partition functions are related
\begin{align}\label{ZF2}
\begin{split}
Z_{{\cal F}}[AE]=&\frac{1}{2}(Z_{{\cal F}}[AP]+Z_{{\cal F}}[AA]), \\
Z_{{\cal F}}[AO]=&-\frac{1}{2}(Z_{{\cal F}}[AP]-Z_{{\cal F}}[AA]),\\
Z_{{\cal F}}[PE]=&\frac{1}{2}(Z_{{\cal F}}[PP]+Z_{{\cal F}}[PA]), \\
Z_{{\cal F}}[PO]=&-\frac{1}{2}(Z_{{\cal F}}[PP]-Z_{{\cal F}}[PA])\,.
\end{split}
\end{align}
The boundary conditions $AP,PA,AA$ correspond to the even spin structures and
$PP$ corresponds to the odd spin structure.
Thus, on a torus, \eqref{FFp} implies
\ie\label{IFTOstack}
\bullet~~&\text{IFTO Stacking:}\\
&Z_{\cal F} [AA]  = Z_{{\cal F}'}[AA]\,,~~~Z_{\cal F} [AP]  = Z_{{\cal F}'}[AP]\,,\\
&Z_{\cal F} [PA]  = Z_{{\cal F}'}[PA]\,,~~~Z_{\cal F} [PP]  =-  Z_{{\cal F}'}[PP]\,.
\fe
This implies that ${\cal F}$ and ${\cal F}'$ share the same Hilbert space in the anti-periodic sector, while the
fermion parity differs by a sign in the periodic sector. 

If a fermionic CFT ${\cal F}_0$ satisfies
\ie\label{F0}
Z_{{\cal F}_0 }  [\rho ]=  Z_{\rm IFTO}[\rho]Z_{{\cal F}_0}  [\rho ]\,.
\fe
then  ${\cal F}_0$ remains unchanged after
stacking with an IFTO .  This means that the  fermionic CFT ${\cal F}_0$ is at
the phase transition boundary between a trivial order and an IFTO.  In other
words, the fermionic CFT ${\cal F}_0$ describes a continuous phase transition 
between two fermionic CFTs $\cal F$ and $\cal F'$ that differ by an IFTO.  For such a fermionic CFT, it satisfies
\ie\label{IFTOinv}
Z_{ {\cal F}_0} [\rho]= 0\,,~~~\rho:~\text{odd}\,.
\fe
On the torus, its partition function with periodic conditions on both cycles ($PP$)
vanishes, {\it i.e.} $Z_{ {\cal F}_0} [PP]=0$. 

The classic example is a single Majorana fermion theory, governed by the action 
\begin{align}
S=\frac{1}{2\pi}\int d^2 z \left(\psi \bar{\partial}\psi+\bar{\psi}\partial\bar\psi +\ii m\bar\psi \psi\right)\,,
\end{align} where $\psi (z)$ and $\bar{\psi} (\bar z)$ are right and left-moving Majorana fermion field. It is the low energy description of Kitaev's fermionic chain model \cite{Kitaev:2001kla} near the transition from the trivial insulator to the $p$-wave superconductor. Here ${\cal F}[m]$ is the
Majorana fermion with mass $m$.  The difference of a Majorana fermion with
positive mass $m$ and that with a negative mass $-m$ is precisely the IFTO:
\ie
Z_{\rm Maj}[\rho, m] = Z_{\rm IFTO}[\rho] \, Z_{\rm Maj}[\rho, -m]\,.
\fe
This implies at the critical point $m=0$, a single massless Majorana fermion CFT
${\cal F}_0={\cal F}[0]$, is  invariant under the stacking with the IFTO.

\paragraph{Exactly Marginal Deformation}
In the example of  a single Majorana fermion, the mass term, {\it i.e.}
the deformation operator $\cal O$ that drives the transition through ${\cal F}_0$, is relevant. 
When $\cal O$ is  exactly  marginal, it moves the CFT $\mathcal{F}_0$ along two different directions onto  the
conformal manifold (leaving the central charge unchanged). 
 Let $g$ be the exactly marginal coupling, we will denote the fermionic CFT on the conformal manifold as ${\cal F}[g]$ with ${\cal F}[g=0]={\cal F}_0$. 
The main point of this paper is to give examples of 
the  transition on a one-dimensional locus of the conformal manifold, where
CFTs on the two sides ${\cal F}[g]$ and ${\cal F}[-g]$ differ by an IFTO, much as the
positive mass and the negative mass Majorana fermions do. 
More precisely,
\ie
Z_{{\cal F} }[\rho, g] = Z_{\rm IFTO}[\rho] \, Z_{{\cal F} }[\rho, -g]\,,
\fe
point-by-point for every $g$.\footnote{This difference in the IFTO can also be thought of as an anomaly involving the  coupling $g$ and the spin structure $\rho$ in the sense of \cite{Cordova:2019jnf,Cordova:2019uob}.} 
We will call such a transition  ${\cal F}[-g]\to {\cal F}_0 \to {\cal F}[g]$ on the conformal manifold a \textit{topological transition}.  
This is to be distinguished from the standard second order phase transition where the gap is closed at the critical point but then opens again. 
In the topological transition, every point is a CFT and the gap is always zero. 

More generally,  starting from a fermionic CFT ${\cal F}_0$, we can turn on two different exactly marginal deformations ${\cal O}$ and ${\cal O}'$ with couplings $g\ge0$ and $g'\ge0$.  
There can also be topological transitions from ${\cal F}'[g']\to {\cal F}_0\to {\cal F}[g]$ such that ${\cal F}[g]$ differs from ${\cal F}'[f(g)]$ by  an IFTO for some function $f(g)$.

\begin{figure*}[th]
\includegraphics[scale=0.9]{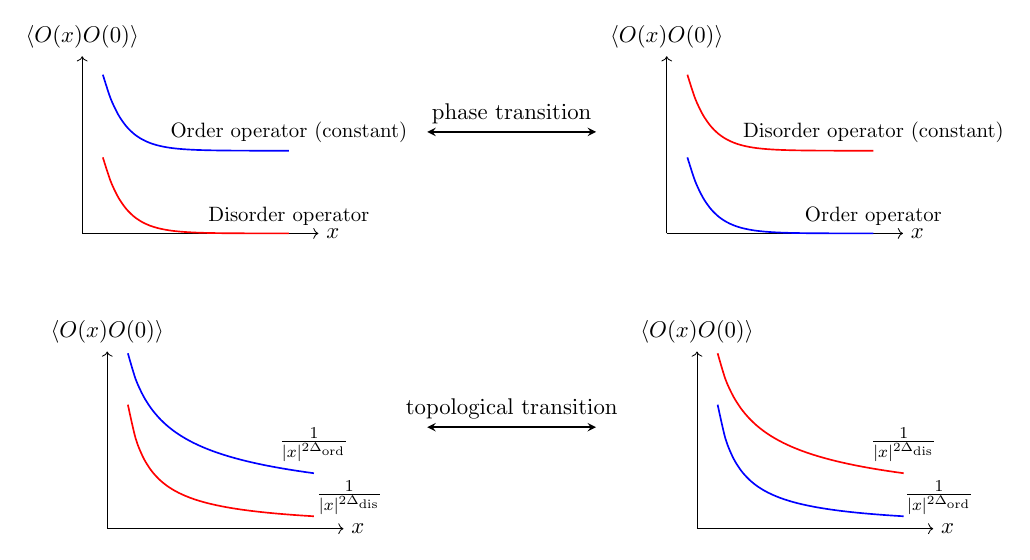}
\caption{Top: The Landau phase transition between the symmetry breaking phase (the order phase) and the symmetric phase (the disorder phase). Bottom: The topological transition between two families of gapless phases with  power-law decaying correlation functions for both the order and the disorder operators. In one family, the scaling dimension of the order operator is smaller than that of the disorder operator, {\it i.e.} $\Delta_{\rm ord}<\Delta_{\rm dis}$, while in the other family we have $\Delta_{\rm ord}>\Delta_{\rm dis}$.}\label{fig:toptrans}
\end{figure*}

In the order phase of a standard second order phase transition,  the two-point function of an order operator approaches a constant at large separation, while that of the disorder operator decays exponentially.  The situation is reversed  in the disorder phase.  
By contrast, along the topological transition ${\cal F}'\to {\cal F}_0 \to {\cal F}$ on the conformal manifold, the two-point functions of the order and disorder operators both fall off by power laws. 
We illustrate the two-point functions of order and disorder operators in both cases in Figure \ref{fig:toptrans}. 
The critical exponents ({\it i.e.} the scaling dimensions) of the  order operator and the disorder operator cross over with each other at the transition point ${\cal F}_0$. 
We will demonstrate this in explicit examples in Section \ref{sec:Fmodels}. 

The mapping from ${\cal F}$ to ${\cal F}'$, {\it i.e.} the stacking with the
IFTO, resembles a $\bZ_2$ transformation. Indeed, in the Majorana fermion
example, such a $\bZ_2$ transformation is the chiral fermion parity
$(-1)^{F_L}$, which flips the sign of the left-moving fermion but not that of
the right.
 This $\bZ_2$ transformation maps ${\cal F}[m]$ to ${\cal
F}[-m]$, and it is a global symmetry of the theory at the transition point ${\cal F}_0$.  
More generally in the topological transition, we will denote this $\bZ_2$ action as $\bZ_2^{\rm IFTO}$, since its action is to stack an IFTO.\footnote{Despite the notation might have suggested, $\bZ_2^{\rm IFTO}$ is \textit{not} a global symmetry of the IFTO \eqref{FSPT} (but of the theory ${\cal F}_0$). The only $\bZ_2$ symmetry of the IFTO is the fermion parity $(-1)^F$.}

\paragraph{(2+1)$d$ $\bZ_2$ Topological Order}

The 't Hooft anomaly of a $\bZ_2$ internal, unitary global symmetry in (1+1)$d$ has a $\bZ_8$ classification   \cite{PhysRevB.85.245132,Qi_2013,PhysRevB.88.064507,PhysRevB.89.201113,Kapustin:2014dxa,FH160406527}.  
The $\bZ_2^{\rm IFTO}$ has odd units of the mod 8 anomaly (see Section \ref{app:ext}). 
Consequently, there is no (1+1)$d$ lattice UV completion of
${\cal F}_0$ such that $\bZ_2^\text{IFTO}$ is realized as an on-site symmetry.  
Instead, there is a (2+1)$d$ lattice UV completion of ${\cal F}_0$ as a boundary theory,
such that $\bZ_2^\text{IFTO}$ is realized as an on-site symmetry on the (2+1)$d$
lattice \cite{HL160607816,cheng2017exactly}.  To see this, we note that the four-component partition functions for a
fermionic CFT ${\cal F}_0$ is given by the partition functions on the four sectors
of boundary of the (2+1)$d$ $\bZ_2$ topological order  ({\it
i.e.} the untwisted $\bZ_2$ gauge theory \cite{Dijkgraaf:1989pz}):
\begin{align}
\label{ZfZ}
\begin{split}
&Z_{{\cal F}}[AE] =Z_1, \ \ \ Z_{{\cal F}}[PO] =Z_e, \\
&Z_{{\cal F}}[PE] =Z_m, \ \ \ Z_{{\cal F}}[AO] =Z_f,
\end{split}
\end{align}
that are labeled by four types of anyons: $1,e,m,f$ \cite{JW190513279}.  
Physically, the operators in each of the four sectors  ${\cal H}_{A/P}^{E/O}$ become the operators on the (1+1)$d$ boundary that  live at the end of the corresponding anyon in the coupled system. 
The $\bZ_2^\text{IFTO}$ transformation maps $Z_{\cal{F}}[PP]$ to $-Z_{\cal{F}} [PP]$ for the boundary theory on a torus. Thus, in the (2+1)$d$ system, the IFTO-stacking
$\bZ_2^\text{IFTO}$ symmetry is  the  electromagnetic duality (which is a 0-form $\bZ_2$ symmetry of the $\bZ_2$ gauge theory) that exchanges $e$ and $m$
but leaves $1$ and $f$ unchanged.

\section{Invertible Fermionic Topological Order and $\bZ_2$ Orbifold}\label{sec:bosferm}

In this section we provide an equivalent bosonic description of the topological transition for fermionic CFT discussed in Section \ref{sec:phase}.

\subsection{Bosonization and Fermionization}\label{sec:BF}

We start by reviewing the procedure of bosonization and fermionization in (1+1) dimensions that has been developed in \cite{Gaiotto:2015zta,Kapustin:2017jrc,Thorngren:2018bhj,Karch:2019lnn,yujitasi}.  See \cite{Radicevic:2018okd} for related discussions on the lattice from a modern perspective.

\paragraph{Fermion$\to$ Boson}

Our starting point is a general (1+1)-dimensional fermionic QFT $\mathcal{F}$ with partition function $Z_{\cal F} [\rho] $.  
A universal symmetry for any fermionic QFT is the fermion parity $(-1)^F$.  The partition function with a nontrivial background field $S$ for the $(-1)^F$ is $Z_{\cal F}[S+\rho]$. 

Next, we would like to gauge the $(-1)^F$ to obtain a bosonic theory $\cal B$ which is independent of the choice of the spin structure.  
We will promote the $(-1)^F$ background gauge field $S$ to a dynamical gauge field $s$, and sum over it with an overall normalization factor ${1\over 2^g}$.
The resulting partition function of the bosonic theory is
\ie
Z_{\cal B}=  {1\over 2^g} \sum_s  Z_{\cal F}[s+\rho] \,.
\fe
Note that the RHS is independent of the choice of $\rho$.

In (1+1) dimensions, gauging a $\bZ_2$ symmetry (in this case, the fermion parity $(-1)^F$) gives rise to a dual $\bZ_2^{\cal B}$ symmetry \cite{Vafa:1989ih} in the gauged theory.  
The partition function of the bosonic theory $\cal B$ with a nontrivial dual $\bZ_2^{\cal B}$ background field $T$ is
\ie\label{ZBT}
Z_{\cal B}[T]  = & {1\over 2^g} \sum_s  Z_{\cal F}[s+\rho]  \nonumber\\
&\cdot \exp\left[ \ii \pi  \left(
\int s\cup T  + {\rm Arf}[T+\rho] +{\rm Arf}[\rho]\right)\right]\,.
\fe
Indeed,  one can check that the RHS is independent of the choice of the spin structure $\rho$.  
We will call the bosonic theory $\cal B$ the \textit{bosonization} of the fermionic theory $\cal F$.  
Note that in this terminology, bosonization is a map from a fermionic theory to a bosonic theory, not an equivalence of the two theories.  
In string theory, this is known as the   Gliozzi-Scherk-Olive projection \cite{Gliozzi:1976qd}.

\paragraph{Boson $\to$ Fermion}

Suppose instead we start with a bosonic theory $\cal B$ with a non-anomalous $\bZ_2^{\cal B}$ symmetry. 
How do we obtain a fermionic theory via gauging?  
We first couple $\cal B$ to the $\bZ_2^{\cal B}$ background field $T$ in a way that depends on the choice of the spin structure:
\ie\label{ZBT}
Z_{\cal B}[T]\, e^{\ii\pi {\rm Arf}[T+\rho] +\ii \pi {\rm Arf}[\rho]}\,.
\fe

This can be interpreted as coupling the bosonic CFT $\cal B$ to the  IFTO via the term ${\rm Arf}[T+\rho]$.  
Next, we promote the background field $T$ to a dynamical field $t$, and obtain a fermionic theory that depends  on the spin structure,\footnote{For those not familiar about cup products, The condensed-matter oriented reference are \cite{Chen:2011pg} }
\ie\label{ZFS}
Z_{\cal F}[S+\rho ] =& {1\over 2^g} \sum_t Z_{\cal B}[t] \nonumber\\
&\cdot \exp\left[
\ii \pi
\left(
 {\rm Arf}[t+\rho] + {\rm Arf}[\rho] +\int t\cup S
 \right)
 \right]\,.
\fe
In the resulting fermionic theory, the dual symmetry is identified as the fermion parity $(-1)^F$, and $S$ is its background field. 
We will call the fermionic theory $\cal F$ the \textit{fermionization} of the bosonic theory $\cal B$ with respect to $\bZ_2^{\cal B}$.\footnote{In the context of vertex operator algebra, the fermionization $\cal F$ is called a non-local $\bZ_2^{\cal B}$ cover of $\cal B$  \cite{Dixon:1988qd}.} 
This can be thought of as the continuum version of the  Jordan-Wigner transformation on the lattice \cite{Jordan:1928wi}. 
Using  \eqref{complete}, we see that the fermionization \eqref{ZFS} with respect to the  $\bZ_2^{\cal B}$ symmetry is the inverse of bosonization (i.e.\ gauging $(-1)^F$)  \eqref{ZBT}.  
Importantly, the fermionization depends on a choice of a non-anomalous $\bZ_2^{\cal B}$ global symmetry.\footnote{The fermionization described above does not hold when the $\bZ_2^{\cal B}$ is anomalous, in which case the partition function $Z[A]$ depends not just on the cohomology class of the background field $A$, but also on the choice of the representative.}    
Generally, a bosonic theory might have more than one non-anomalous $\bZ_2$ symmetries, and its fermionization might not be unique.

\paragraph{Torus Partition Functions}

Let us apply the bosonization and fermionization procedures to torus partition functions of CFTs. 
In a bosonic CFT with a global symmetry $\bZ_2^{\cal B}$, we define $Z_{{\cal B}}[\alpha_x \alpha_t]$ with $\alpha_x,\alpha_t=0,1$ as the torus partition
functions with ($\alpha=1$) or without ($\alpha=0$) the $\bZ_2^{\cal B}$ twists  in the $x$ and $t$
directions. 
These torus partition functions have the following trace interpretations:
\ie
&Z_{\cal B} [00]  = \text{Tr}_{\cal H}\left[  \,     \,q^{ H+K\over2 } \bar q^{ H-K\over2 }\, \right]\,,\\
&Z_{\cal B} [01]  = \text{Tr}_{\cal H}\left[ \, \eta   \,q^{ H+K\over2 } \bar q^{ H-K\over2 }  \,\right ]\,,\\
&Z_{\cal B} [10]  = \text{Tr}_{\widetilde{\cal H}}\left[  \,   \,q^{ H+K\over2 } \bar q^{ H-K\over2 }  \,\right] \,,\\
&Z_{\cal B} [11]  = \text{Tr}_{\widetilde{\cal H}}\left[ \, \eta   \,q^{ H+K\over2 } \bar q^{ H-K\over2 } \,\right] \,,
\fe
where $\cal H$ is the Hilbert space where all operators have the periodic boundary condition, while $\widetilde{\cal H}$ is the Hilbert space where the $\bZ_2^{\cal B}$-even ($\bZ_2^{\cal B}$-odd) operators have the periodic (antiperiodic) boundary condition.  
Here $\eta$ is the $\bZ_2^{\cal B}$  charge operator. 
We will call $\cal H$ the \textit{untwisted sector} and $\widetilde {\cal H}$ the \textit{twisted sector} with respect to $\bZ_2^{\cal B}$.  
Via the operator-state correspondence, states in $\cal H$ are in one-to-one correspondence with the local operators, while states in $\widetilde {\cal H}$ are in one-to-one correspondence with the non-local operators living at the end of the $\bZ_2^{\cal B}$ line defect \cite{Chang:2018iay}.

Alternatively, we can consider the $\bZ_2$-even/odd subsectors ${\cal H}^{E/O}$ of the untwisted sector $\cal H$, and similarly the $\bZ_2^{\cal B}$-even/odd subsectors $\widetilde {\cal H}^{E/O}$ of the twisted sector $\widetilde{\cal H}$. The associated torus partition functions are
\begin{align}
\begin{split}
Z_{{\cal B}}[0E]=&\frac{1}{2}(Z_{{\cal B}}[00]+Z_{{\cal B}}[01])\,,\\
Z_{{\cal B}}[0O]=&\frac{1}{2}(Z_{{\cal B}}[00]-Z_{{\cal B}}[01])\,,\\
Z_{{\cal B}}[1E]=&\frac{1}{2}(Z_{{\cal B}}[10]+Z_{{\cal B}}[11])\,,\\
Z_{{\cal B}}[1O]=&\frac{1}{2}(Z_{{\cal B}}[10]-Z_{{\cal B}}[11]).
\end{split}
\end{align}

Following the fermionization procedure \eqref{ZFS}, we can relate the four bosonic torus partition functions $Z_{\cal B}  [0E]$, $Z_{\cal B}  [0O]$, $Z_{\cal B}  [1E]$, $Z_{\cal B}  [1O]$ to the four fermionic torus partition functions $Z_{\cal F}[ AE],Z_{\cal F}[ AO], Z_{\cal F}[ P E],Z_{\cal F}[ P O]$ (defined in \eqref{ZF1} and \eqref{ZF2}):
\ie\label{torusBF}
\bullet~~&\text{Fermionization/Bosonization}:\\
&Z_{\cal F} [AE]  = Z_{\cal B}  [0E]\,,~~~~Z_{\cal F} [AO ] =  Z_{\cal B} [1 O]\,,\\
&Z_{\cal F}[PE]  = Z_{\cal B}[0O]\,,~~~~Z_{\cal F} [PO]  =Z_{\cal B}[1E]\,.  
\fe
 
As a special example, consider the case when $\cal F$ is the massless Majorana fermion and $\cal B$ is the Ising CFT.  
 The  Virasoro primaries and their conformal weights $(h,\bar h)$ in the four Hilbert spaces of the Majorana fermion CFT, or equivalently via \eqref{torusBF},  the four sectors of the Ising CFT Hilbert space, are
 \ie\label{Isingtable}
&   {\cal H}_A^E[\text{Maj}] = {\cal H}^E[\text{Ising}] ~~~1:(0,0)\,,~~~ \varepsilon :(\frac 12,\frac 12)\,,\\
 &  {\cal H}_A^O[\text{Maj}] =\widetilde{\cal H}^O[\text{Ising}] ~~~\psi :(\frac12, 0)\,,~~~\bar\psi :(0,\frac 12)\,,\\
 &  {\cal H}_P^E[\text{Maj}] = {\cal H}^O[\text{Ising}]~~~ \sigma:({1\over 16},{1\over16})\,,\\
 & {\cal H}_P^O[\text{Maj}]= \widetilde{\cal H}^E[\text{Ising}] ~~~\mu:({1\over 16},{1\over 16})\,.
 \fe
Here $\psi,\bar\psi$ are the left- and the right-moving Majorana fermions,  $\varepsilon = \psi\bar\psi$ is the energy operator, $\sigma$ is the spin (or order) operator, and $\mu$ is the disorder operator.

This  bosonization/fermionization relation generalizes the familiar relation between the Ising CFT and the Majorana fermion  to any bosonic CFT with a non-anomalous $\bZ_2^{\cal B}$ global symmetry and any fermionic CFT.  
In going from $\cal F$ to $\cal B$, we  gauge the $ (-1)^F$ of $\cal F$ to obtain a bosonic theory $\cal B$ with $\bZ_2^{\cal B}$ symmetry via \eqref{ZBT}. 
Conversely, using \eqref{ZFS}, we gauge $\bZ_2^{\cal B}$ (with a nontrivial coupling to the IFTO) of $\cal B$ to retrieve the fermionic theory $\cal F$ we start with.

\subsection{Gauging with the Invertible Fermionic Topological Order}\label{sec:gaugeIFTO}

Starting with a fermionic theory $\cal F$, consider another fermionic theory $\cal F'$ defined as multiplying $\cal F$ by the IFTO \eqref{FSPT}:
\ie
Z_{\cal F'}[S'+\rho]  =Z_{\cal F} [S'+\rho ] \,e^{ \ii \pi {\rm Arf}[S'+\rho]}\,.
\fe
Here $S'$ is the background field for the $(-1)^F$ symmetry in $\cal F'$.  
The partition functions for ${\cal F}$ and ${\cal F}'$ are identical  on Riemann surfaces with even spin structures, but differ by a sign on manifolds with  odd spin structures.   
Next we gauge the $(-1)^F$ of $\cal F'$ to obtain a bosonic theory $\cal B'$ following the recipe \eqref{ZBT}:
\begin{align}
 &Z_{\cal B'} [T' ]\nonumber\\
=&{1\over 2^g}  \sum_{s'}  Z_{\cal F} [s'+ \rho ] \nonumber \\
&\cdot \exp\left[ \ii \pi  \left(
{\rm Arf}[s'+\rho]+
\int s'\cup T'  + {\rm Arf}[T'+ \rho] +{\rm Arf}[\rho]\right)\right]
\end{align}
We can use \eqref{arfid1} to rewrite the partition function of ${\cal B}'$ as
\ie\label{tempBp}
Z_{\cal B'} [T' ]   ={1\over 2^g}  \sum_{s'}  Z_{\cal F} [s'+ \rho ] \, 
\exp\left[ \ii \pi  
{\rm Arf}[s'+T'+ \rho]\right]
\fe

How is $\cal B'$ related to $\cal B$? Let us consider the $\ZZ_2^{\cal B}$ orbifold of $\cal B$, which has a dual $\ZZ_2$ symmetry. Its partition function for $\cal B'$, with the background field $T'$ for the dual $\ZZ_2$ symmetry is ${1\over 2^g} \sum_t Z_{\cal B}[t]  e^{\ii \pi \int t\cup T'}$, where $t$ is the dynamical gauge field for $\ZZ_2^{\cal B}$. 
It follows from (\ref{ZBT}) that we can write $Z_{{\cal B}'}$ in terms of the fermionic theory $Z_{\cal F}$:
\begin{widetext}
\begin{align}\label{BZ2}
{1\over 2^g} \sum_t Z_{\cal B}[t]  e^{\ii \pi \int t\cup T'} 
&= {1\over 2^g} \sum_t
{1\over 2^g}  \sum_{s}  Z_{\cal F} [s+\rho ] \, 
\exp\left[ \ii \pi  \left(
\int t\cup T'+
\int s\cup t + {\rm Arf}[t+\rho] +{\rm Arf}[\rho]\right)\right]\notag\\
&= 
{1\over 2^g}  \sum_{s'}  Z_{\cal F} [s'+ \rho ] \, 
\exp\left[ \ii \pi  
{\rm Arf}[s'+T'+ \rho] \right] 
\end{align}
\end{widetext}
where in the second line we have used \eqref{arfid2} and renamed $s$ as $s'$. 
Matching \eqref{tempBp} with \eqref{BZ2}, we have shown that $\cal B'$ is the $\bZ_2^{\cal B}$ orbifold of $\cal B$, i.e. ${\cal B}' = {\cal B}/\bZ_2^{\cal B}$.

\paragraph{(2+1)$d$ $\bZ_2$ Topological Order and the Electromagnetic Duality}
The torus partition functions of $\cal B$ and ${\cal B}'$ are related as follows:
\begin{align}\label{Z2orbifold}
\bullet~~&\bZ_2^{\cal B}\text{ orbifold}:\nonumber\\
\begin{split}
& Z_{\cal B}[ 0 E]  = Z_{{\cal B}'} [0E]\,,~~~~Z_{\cal B} [0O ] = Z_{{\cal B}'}[1E]\,,\\
&Z_{\cal B} [ 1E ] = Z_{{\cal B}'} [0O]\,,~~~~Z_{\cal B}[1O]  = Z_{ {\cal B}'} [1O]\,.
\end{split}
\end{align}
The $\bZ_2^{\cal B}$-odd, untwisted sector ${\cal H}^O$ and the $\bZ_2^{\cal B}$-even, twisted sector $\widetilde{\cal H}^E$ are exchanged under the $\bZ_2^{\cal B}$ orbifold. 
The $\bZ_2^{\cal B}$ orbifold in (1+1)$d$ has a natural interpretation from the (2+1)$d$ $\bZ_2$ gauge theory, which we explain below. 
In Section \ref{sec:phase}, we realize the four-component fermionic partition functions in terms
of the four boundary partition functions $Z_{1,e,m,f}$ of the (2+1)$d$ $\bZ_2$
topological order, as in (\ref{ZfZ}).  
 The same four boundary
partition functions $Z_{1,e,m,f}$ also give rise to the  four-component partition
functions for a bosonic CFT with $\bZ_2^{\cal B}$ symmetry \cite{Lin:2019kpn,JW190513279}:
\ie
\begin{split}
&Z_{\cal B}[0E] =Z_1 \,,~~~~Z_{\cal B}[0O]=Z_e\,,\\
&Z_{\cal B}[1E] = Z_m\,,~~~~Z_{\cal B}[1O]=Z_f\,.
\end{split}
\fe
The operators in each of the four sectors ${\cal H}^{E/O}, \widetilde{\cal H}^{E/O}$ become the  operators on the (1+1)$d$ boundary that live at the end of the corresponding anyon in the coupled system. 
From the point of view of the (2+1)$d$ $\bZ_2$ topological order, the $\bZ_2^{\cal B}$ orbifold exchanges $Z_e$ and $Z_m$, which is the electromagnetic duality \cite{rey1991self,Severa:2002qe,Freed:2018cec}. 
Therefore,  the electromagnetic duality of the (2+1)$d$ $\bZ_2$ topological order induces the Kramers-Wannier duality of the boundary (1+1)$d$ bosonic CFTs.

\begin{figure}[th]
\centering
\includegraphics[scale=1]{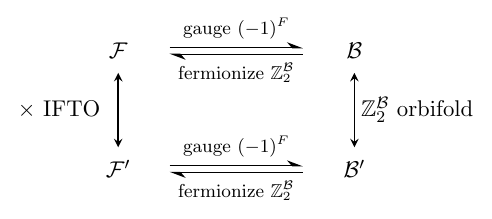}
\caption{The commutative diagram of fermionic CFTs $\cal F$, $\mathcal{F}'$ and their bosonizations $\cal B$ and $\mathcal{B}'$.}\label{fig:CD}
\end{figure}

We summarize the above discussion in the commutative diagram \cite{Kapustin:2014gua,yujitasi} shown in Figure \ref{fig:CD}.
Given any bosonic CFT $\cal B$ with a non-anomalous $\bZ_2^B$ symmetry,  we  obtain two fermionic CFTs $\cal F$ and ${\cal F}'$ that differ by an IFTO. Conversely, given any fermionic CFT $\cal F$ and its IFTO-stacked theory ${\cal F}'$, we  obtain two bosonic CFTs $\cal B$ and ${\cal B}'$ related by a $\bZ_2^{\cal B}$ orbifold.  
Under the IFTO-stacking,  the bosonization/fermionization, and the $\bZ_2^{\cal B}$ orbifold,  the Hilbert spaces are permuted as in \eqref{IFTOstack}, \eqref{torusBF}, and \eqref{Z2orbifold}, respectively. 
Both the fermionic CFT ${\cal F}$ and the bosonic CFT ${\cal B}$ can be realized as the boundary of the (2+1)$d$ untwisted $\bZ_2$ gauge theory. For the fermionic boundary, the electromagnetic duality in the bulk exchanging the electric $e$ and the magnetic $m$ anyons induces the IFTO stacking, while for the bosonic boundary, it induces the $\bZ_2^{\cal B}$ orbifold ({\it i.e.} the Kramers-Wannier duality).

%

\section{Transition between Bosonic CFTs and their $\bZ_2$ Orbifolds}\label{sec:Btransition}

In Section \ref{sec:bosferm}, we see that the IFTO stacking of fermionic CFTs is mapped to the $\bZ_2^{\cal B}$ orbifold of bosonic CFTs via bosonization.  
Therefore, the topological transition introduced in Section \ref{sec:phase} on the fermionic conformal manifold is equivalent to a one-dimensional bosonic conformal manifold where the CFTs are related by a $\bZ_2^{\cal B}$ orbifold point-by-point across the transition point ${\cal B}_0$.  
In this section we discuss several such bosonic examples, paving the way for their corresponding fermionic models in Section \ref{sec:Fmodels}.

\subsection{Kramers-Wannier Duality Defect}\label{subsec:KW}

\paragraph{Kramers-Wannier Duality} 
Let ${\cal B}'\to {\cal B}_0\to {\cal B}$ be the bosonization ({\it i.e.} gauging $(-1)^F$) of the topological transition ${\cal F}'\to {\cal F}_0\to {\cal F}$, and let $\bZ_2^{\cal B}$ be the emergent symmetry from gauging $ (-1)^F$.  
The exactly marginal deformation that interpolates between ${\cal F}$ and ${\cal F}'$ survives the bosonization, and gives an exactly marginal deformation interpolating between $\cal B$ and $\cal B'$. 
From Section \ref{sec:gaugeIFTO} and \eqref{FFp}, we learn that 
\ie
{\cal B}' =  {\cal B}/\bZ_2^{\cal B}\,.
\fe
In particular, at the origin of the deformation, the bosonic CFT ${\cal B}_0$ is self-dual under the $\bZ_2^{\cal B}$ orbifold:
\ie\label{KW}
{\cal B}_0  ={\cal B}_0 /\bZ_2^{\cal B}\,.
\fe
In the example when ${\cal F}_0$ is a single Majorana fermion,  its bosonization ${\cal B}_0$ is the Ising CFT.  
The self-duality \eqref{KW} is then nothing but  the Kramers-Wannier duality \cite{PhysRev.60.252,Monastyrsky:1978kp}. 
More generally, \eqref{KW} generalizes the familiar Kramers-Wannier duality  to any bosonic CFT that is self-dual under $\bZ_2^{\cal B}$ orbifold.

\paragraph{Topological Defect Line}

\begin{figure*}
\centering
\includegraphics[width=.25\textwidth]{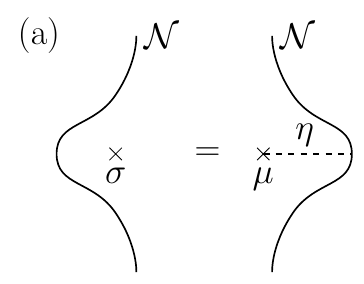}~~
\includegraphics[width=.32\textwidth]{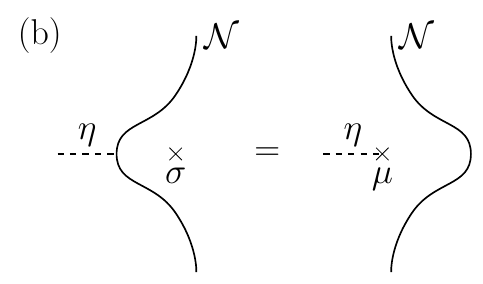}~~
\includegraphics[width=.25\textwidth]{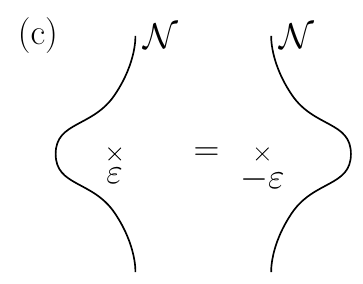}\\
~\\
\centering
\includegraphics[width=.25\textwidth]{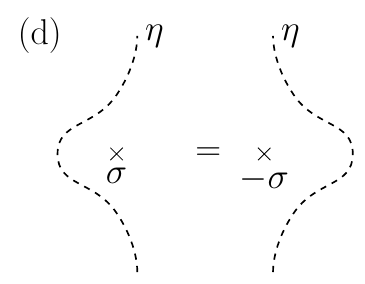}~~~~~~
\includegraphics[width=.25\textwidth]{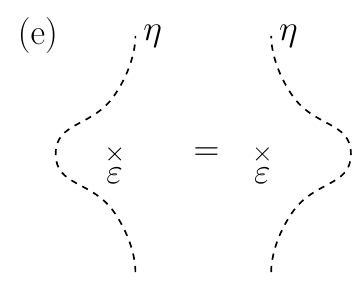}
\caption{The action of topological defect lines on local and non-local operators in the Ising CFT \cite{Frohlich:2004ef}.  The $\bZ_2^{\cal B}$ defect line $\eta$ acts on operators with $\pm1$ sign. The duality defect $\cal N$ exchanges the local, order operator $\sigma(z,\bar z)$ with the non-local, disorder operator $\mu(z,\bar z)$, which lives at the end of the $\bZ_2^{\cal B}$ defect line $\eta$.}\label{fig:defectaction}
\end{figure*}

Our description of the bosonic model will rely on the  {\it topological defect lines}, which are one-dimensional extended objects in spacetime. 
We will give a brief review on this subject, while the readers are referred to \cite{Bhardwaj:2017xup,Chang:2018iay} for a more complete introduction. 
Any global symmetry in (1+1) dimensions is associated with a topological defect line that implements the symmetry action on local operators \cite{Kapustin:2014gua,Gaiotto:2014}.  
However, not all topological defect line is associated with a global symmetry. 
Such topological defect line is called {\it non-invertible}, or non-symmetry \cite{Chang:2018iay}. 
One feature of the non-invertible topological defect line is that its action, when restricted to the {\it local} operators, is not invertible, and therefore not group-like.  
More precisely, as we bring a non-invertible topological defect line past a local operator, we might create a non-local operator.

The Kramers-Wannier duality \eqref{KW} of the bosonic CFT ${\cal B}_0$ is implemented by such a non-invertible  defect $\cal N$, sometimes also called the {\it duality defect} \cite{Frohlich:2004ef,Frohlich:2006ch,Frohlich:2009gb,Bhardwaj:2017xup,Chang:2018iay}.  
The duality defect $\cal N$, together with the $\bZ_2^{\cal B}$ defect $\eta$, form  a fusion category known as the $\bZ_2$ Tambara-Yamagami (TY) category \cite{TAMBARA1998692} with the Ising fusion rules:
\ie\label{TYZ2}
\eta^2 =I \,,~~~~ {\cal N}^2 = 1+\eta\,,~~~~{\cal N}\eta= \eta{\cal  N} ={\cal  N}\,,
\fe
where $I$ is the trivial topological line.\footnote{Given the above fusion rules, there are two solutions to the pentagon identities for the $F$-moves. One of them is realized in the Ising CFT, and other is realized in the ${\mathfrak{su}(2)}_2$ WZW model. See \cite{TAMBARA1998692} for their respective $F$-symbols.}

Let us discuss the  duality defect in the Ising CFT \cite{Frohlich:2004ef} in details as an example.  
In the Ising sector, we will denote the energy operator as $\varepsilon(z,\bar z)$ with conformal weights $(h,\bar h) = (\frac 12,\frac 12)$, and the spin operator (also known as the order operator) as $\sigma(z,\bar z)$ with conformal weights $(h,\bar h)=({1\over 16}, {1\over 16})$.  
  As we sweep the duality defect $\cal N$ past the energy operator $\varepsilon$, the latter obtains a minus sign.  
On the other hand, as we sweep the duality defect $\cal N$ past the spin order $\sigma(z,\bar z)$, a $\bZ_2^{\cal B}$ line $\eta$ is created with the disorder operator $\mu(z,\bar z)$ sitting at the endpoint.  
Therefore the duality defect $\cal N$ exchanges the order operator $\sigma(z,\bar z)$ (which is a local operator) with the disorder operator  $\mu(z,\bar z)$ (which is a non-local operator attached to a line).  
See Figure \ref{fig:defectaction}.

\paragraph{Duality Interface}

\begin{figure*}
\centering
\includegraphics[width=.65\textwidth]{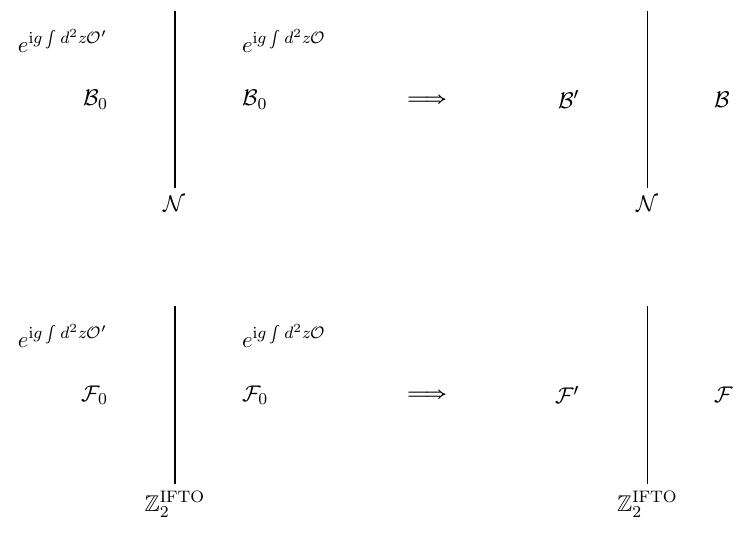}
\caption{The duality defect $\cal N$ in the self-dual bosonic CFT ${\cal B}_0$ becomes a duality interface between   ${\cal B}$ and ${\cal B}'$ after renormalization group flow. The duality defect $\cal N$ in $\cal B$ turns into the $\bZ_2^{\rm IFTO}$  symmetry defect of $\cal F$ under fermionization.}\label{fig:dualityinterface}
\end{figure*}

Now consider  the bosonic CFT ${\cal B}_0$ with a duality defect $\cal N$ inserted along the time direction (see Figure \ref{fig:dualityinterface}). 
Let $\cal O$ be an exactly marginal deformation, and  ${\cal O}'$  be the local operator obtained by sweeping the duality defect $\cal N$ past ${\cal O}$.\footnote{Here we assume ${\cal O}'  \neq\cal O$.  If ${\cal O}'=\cal O$, {\it i.e.} if the duality defect $\cal N$ commutes with $\cal O$, then ${\cal B}= {\cal B}'$ and ${\cal F} ={\cal F}'$, and there is no interesting topological transition to discuss.}  
We now turn on the exactly marginal deformation $\cal O$ to the left of $\cal N$, and ${\cal O}'$ to the right of $\cal N$. 
The deformation drives the systems on the two sides to two different CFTs, ${\cal B}$ and ${\cal B}'$, and the duality defect $\cal N$ becomes a topological {\it duality interface}   between the two CFTs.  
The duality interface implements the duality ${\cal B}' = {\cal B}/\bZ_2^{\cal B}$, but it is not a topological defect line in either $\cal B$ or ${\cal B}'$.

Under fermionization, the duality defect $\cal N$ of the bosonic CFT ${\cal B}_0$ becomes  the $\bZ_2^{\rm IFTO}$ symmetry in the fermionic CFT  ${\cal F}_0$. See Figure \ref{fig:sym} and Section \ref{app:ext} for more discussions.  
By turning on the deformation $\cal O$ and ${\cal O}'$, the $\bZ_2^{\rm IFTO}$ symmetry defect becomes an interface separating two fermionic theories $\cal F$ and ${\cal F}'$ that differ by an IFTO. 
  See Figure \ref{fig:dualityinterface}.

\begin{figure}[h!]
\centering
\includegraphics[width=.5\textwidth]{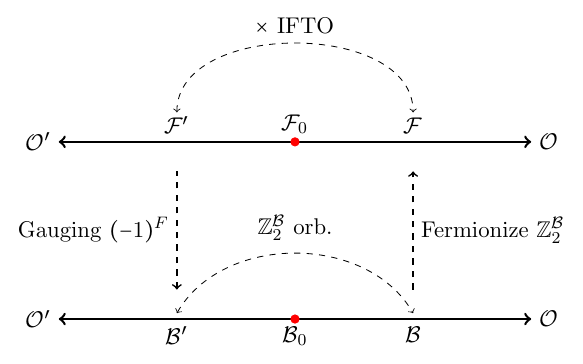}
\caption{The topological transition on the fermionic conformal manifold can be equivalently bosonized to a family of bosonic CFTs that are related by the $\bZ_2^{\cal B}$ orbifold.}\label{fig:bosonfermion}
\end{figure}

  To summarize, the topological transition of fermionic CFT can be equivalently recast into the following bosonic data (see Figure \ref{fig:bosonfermion}): 
\begin{itemize}
\item A  bosonic CFT ${\cal B}_0$ that is self-dual under gauging  a non-anomalous $\bZ_2^{\cal B}$ global symmetry, {\it i.e.} ${\cal B}_0  = {\cal B}_0  /\bZ_2^{\cal B}$.  The self-duality is implemented by  a duality defect $\cal N$.
\item  An exactly marginal deformation ${\cal O}$.
\end{itemize}

\begin{figure*}
\begin{align*}
\left.\begin{array}{ccc}\text{Fermionic CFT}~{\cal F}_0 &  &\text{Bosonic CFT}~ {\cal B}_0 \\
\hline
&&\\
\text{fermion parity}&\xleftrightarrow{\text{~~~~dual~~~~}} &\text{non-anomalous}\\
(-1)^F && \bZ_2^{\cal B} \\
&&\\
\text{IFTO-stacking} & ~~~~~ \xrightarrow{\text{~non-symmetry extension~}}	~~~~~& \text{Duality defect}  \\
\bZ_2^{\rm IFTO} &&{\cal N}
\end{array}\right.
\end{align*}
\caption{Under bosonization/fermionization, the fermion parity $(-1)^F$ is the dual symmetry of $\bZ_2^{\cal B}$, while the IFTO stacking $\bZ_2^{\rm IFTO}$ symmetry of ${\cal F}_0$ is extended to a non-invertible duality defect $\cal N$ in ${\cal B}_0$.}\label{fig:sym}
\end{figure*}

\subsection{Free Compact Boson $S^1$}\label{Sec:FreeBoson}

\begin{figure}[h!]
\centering
\includegraphics[width=.5\textwidth]{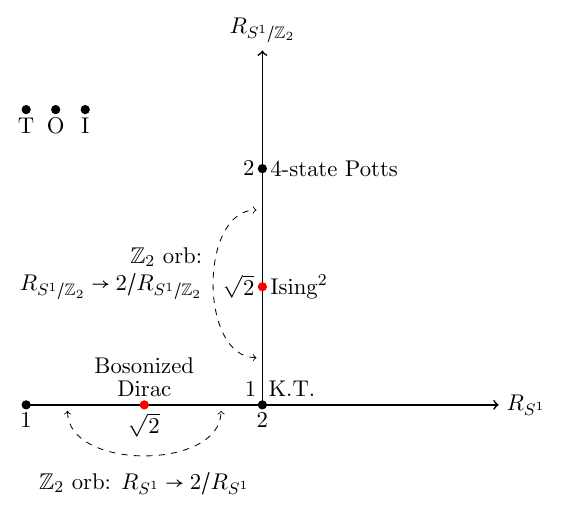}
\caption{The conformal manifold of the bosonic $c=1$ CFTs.  The conformal manifold has an $S^1$ branch labeled by $ R_{S^1}\ge1$, and an orbifold branch $S^1/\bZ_2$ labeled by $ R_{S^1/\bZ_2}\ge1$. The end point of $S^1$ branch at $R_{S^1}=1$ is the $\mathfrak{su}(2)_1$ WZW model. The two branches meet at the Kosterlitz-Thouless point, which is described by $R_{S^1} = 2$ or equivalently by $R_{S^1/\bZ_2}= 1$.  The bosonized Dirac fermion and the Ising$^2$ theories are the self-dual points of the $\bZ_2^{\cal B}$ symmetries defined in \eqref{Z2BS1} and \eqref{Z2BS1Z2}, respectively. The $\mathfrak{su}(2)_1/\Gamma$ orbifold models for $\Gamma=$T, O, I are isolated points in the moduli space. T, O, I each represents the tetrahedral, octahedral and icosahedral groups.}\label{fig:c=1}
\end{figure}

Our first example is the $c=1$ free compact boson theory  (see, for example, \cite{Ginsparg:1988ui} for a review).  
The conformal manifold of $c=1$ CFTs consists of two branches, the $S^1$ branch and the $S^1/\bZ_2$ branch, together with three isolated points. 
See Figure \ref{fig:c=1}. 
In this section we start with the $S^1$ branch, which has the description of the free compact boson $X(z, \bar z) = X_L(z) + X_R(\bar z)$ with identification $X(z,\bar z)\sim X(z,\bar z)+2\pi R$ with  $R \ge 1$. 
Our convention for the radius $R$ is such that T-duality acts as\footnote{Our convention for the radius is related to that in  \cite{Ginsparg:1988ui} as $R^{\rm Ginsparg's}  = R^{\rm Ours} /\sqrt{2}$.}
\ie
\text{T-duality}:~S^1[R] =  S^1\left [ {1 \over R}\right]\,.
\fe
The free boson field is normalized such that $X(z,\bar z) X(0,0 )\sim -\frac 12 \log|z|^2$.  
On the $S^1$ branch, the theory has the $\mathfrak{u}(1)\times \overline{\mathfrak{u}(1)}$  chiral algebra generated by the currents $\partial X(z)$ and $\bar\partial X(\bar z)$.  
At a generic radius, there is one exactly marginal operator generating the conformal manifold:
\ie\label{EMS1Z2}
{\cal O}  =  \partial X \bar\partial X\,.
\fe

\paragraph{Primary Operators}
The local primary operators with respect to the $\mathfrak{u}(1)\times \overline{\mathfrak{u}(1)}$  chiral algebra are
\ie\label{expop}
&V_{n,w}(z,\bar z) \\
=&\exp\left[ \ii \left( {n\over R}+wR\right)  X_L (z)+ \ii \left( {n\over R}- wR\right)  X_R(\bar z) \right],
\fe
which are labeled by two integers, the momentum number $n\in\bZ$ and the winding number $w\in \bZ$. 
The conformal weights of $V_{n,w}$ are
\ie\label{expweight}
h = \frac 14 \left( {n\over R}+wR\right)^2\,,~~~~\bar h=\frac 14\left( {n\over R}-wR\right)^2.
\fe
The torus partition function $Z_{S^1}(R)$ is therefore:\footnote{Here, $\eta (q)$ is the Dedekind eta function defined as $\eta(q ) = q^{1/24}\prod_{i=1}^\infty (1-q^i)$.\label{fn:eta}}
\ie\label{ZS1}
Z_{S^1}(R) =  {1\over |\eta(q)|^2}  \sum_{n,w\in \bZ}  q^{ \frac 14 ( {n\over R} +wR)^2   }\, \bar q^{ \frac 14 ( {n\over R} -wR)^2   }\,.
\fe

The global symmetry at a generic radius contains $( U(1)_n \times U(1)_w ) \rtimes \bZ_2$, where the  $\mathbb{Z}_2$ acts as $X \to -X$. 
The $U(1)_n$ and $U(1)_w$ correspond to momentum and winding, which act by phases $e^{\ii n\theta}$ and $e^{\ii w\theta}$ on the primary operator \eqref{expop}, respectively.  
They act on $X_L(z)$ and $X_R(\bar z)$ by shifts:
\ie
U(1)_n:~~&X_L(z) \to X_L(z)+ \frac{R}{2} \theta_n \,,\\
&X_R(\bar z) \to X_R(\bar z)+  \frac{R}{2} \theta_n \,,\\
U(1)_w:~~&X_L(z) \to X_L(z)+{1\over  2R} \theta_w \,,\\
&X_R(\bar z) \to X_R(\bar z)- {1\over 2R} \theta_w \,,
\fe
with $\theta_{n,w} \sim \theta_{n,w}+2\pi$. 
Both $U(1)_n$ and $U(1)_w$ are non-anomalous for all $R$. 
In particular, this implies that $U(1)_n$ is neither holomorphic nor anti-holomorphic at any radius $R$. 
The same is true for $U(1)_w$. 
 When $R^2$ is rational, certain integral combination  of $U(1)_n$ and $U(1)_w$ becomes holomorphic or anti-holomorphic, and the CFT enjoys an enhanced chiral algebra.

Let $ \bZ_2^{(1,0)}$ and $\bZ_2^{(0,1)}$ be the $\mathbb{Z}_2$ subgroups of $U(1)_n$ and $U(1)_w$, which act on the primary  operators by signs $e^{\ii \pi n}$ and $e^{\ii \pi w}$, respectively.  
There is no 't Hooft anomaly for the momentum $\bZ_2^{(1,0)}$, nor for the winding $\bZ_2^{(0,1)}$ alone, but there is a mixed anomaly between the momentum $\bZ_2^{(1,0)}$ and the winding $\bZ_2^{(0,1)}$.

\paragraph{Twisted Sector $\widetilde{\cal H}$} 
Let us discuss the non-local operators that live in the twisted sector with respect to a $\bZ_2$ global symmetry.    
The twisted sector operators of $\bZ_2^{(m_1,m_2)}$ $(m_i=0,1)$ are given by the same form as \eqref{expop}, but generally with fractional momentum $\tilde n$ and winding number $\tilde w$:
\ie
V_{\tilde n, \tilde w}(z,\bar z) :~~~\tilde n \in {m_2 \over2} +\bZ\,,~~~\tilde w \in {m_1 \over2} +\bZ\,.
\fe
In other words, the momentum $\bZ_2^{(1,0)}$ twist makes the winding number fractional due to the mixed anomaly, and vice versa.

\paragraph{$\bZ_2^{\cal B}$ Orbifold of $S^1$}

We will choose the $\bZ_2^{\cal B}$ symmetry to be the momentum $\bZ_2^{(1,0)}$:
\ie\label{Z2BS1}
\bZ_2^{\cal B}:~V_{n,w} \to (-1)^n V_{n,w}\,.
\fe
  The $\bZ_2^{\cal B}$ orbifold of the $c=1$ compact boson theory at radius $R$ is another compact boson at radius $R/2$, which by T-duality is equivalent to the theory at radius $2\over R$:
\ie\label{KWS1}
{S^1[R]\over \bZ_2^{\cal B} } = S^1\left[{2\over R}\right]\,.
\fe
We compute the twisted torus partition functions of the $S^1[R]$ theory and give  the proof of this relation in Appendix \ref{app:ZS1}. 

The fixed point of this orbifold, {\it i.e.} the Kramers-Wannier self-dual point, is at $R=\sqrt{2}$, which is described by the bosonization of a Dirac fermion. 
This is our first example of a family of bosonic CFTs related by the $\bZ_2^{\cal B}$ orbifold. 

The Kramers-Wannier duality is implemented by  the (0+1)$d$ duality defect line $\cal N$ satisfying \eqref{Z2BS1}. 
As we bring a local operator $V_{n,w}$ past through $\cal N$, it is mapped to (see Section \ref{subsec:Dirac} for derivation)
\begin{align}
 V_{n,w}\rightarrow V_{-2w,-\frac{n}{2}}\,.
\end{align}
The righthand side is only a local operator if $n \in 2\bZ$, {\it i.e.} when ${V}_{n,w}$ is $\bZ_2^{\cal B}$ even.  
On the other hand, when ${V}_{n,w}$ is $\bZ_2^{\cal B}$ odd, {\it i.e.} $n\in 2\bZ+1$, it is mapped to a {\it non-local} operator living at the end of the $\bZ_2^{\cal B}$  line $\eta$. 
This is indeed the characteristic way how a duality defect acts on operators. 
For example, the duality defect $\cal N$ in the Ising CFT maps the local, $\bZ_2^{\cal B}$-odd, order operator $\sigma$  to the non-local, disorder operator $\mu$, while it maps the local, $\bZ_2^{\cal B}$-even, energy operator $\varepsilon$ to itself with a sign. 
See  \cite{fuchs2007topological}  for a thorough discussion on the topological defect lines in $c=1$ CFTs.

\subsection{Bosonic $S^1/\bZ_2$ Orbifold}\label{sec:S1Z2}

The next example is the $c=1$  theory $S^1/\bZ_2[R]$  defined as the $\bZ_2$ orbifold of the $c=1$ compact free boson theory at radius $R$, where the $\bZ_2$ acts as $X\to -X$.  The exactly marginal operator is again \eqref{EMS1Z2}.

The torus partition function of the $c=1$ $S^1/\bZ_2$ orbifold theory at radius $R$  is\footnote{Here $\theta_2(q) = 2q^{1/8}\prod_{i=1}^\infty (1-q^i)(1+q^i)^2$, $\theta_3(q) = \prod_{i=1}^\infty (1-q^i)(1+q^{i-1/2})^2$, $\theta_4(q) = \prod_{i=1}^\infty (1-q^i)(1-q^{i-1/2})^2$, and $\eta (q)$ is the Dedekind eta function as given in footnote \ref{fn:eta}.
}
\ie\label{Zorb}
Z_{S^1/\bZ_2} (R) = \frac 12 Z_{S^1}(R)  + \left| {\eta(q)\over \theta_2(q)} \right|
+ \left| {\eta(q)\over \theta_4(q) }\right|+ \left| {\eta(q)\over \theta_3(q)} \right|\,,
\fe
where the first two terms come from  the untwisted sector and the last two terms come from the twisted sectors of the $S^1$ theory.  
The latter comes from the two twist fields $\sigma_{1,2}(z,\bar z)$, corresponding to the two fixed points of $S^1/\bZ_2$ and their descendants. 
Both $\sigma_{1,2}$ have $h=\bar h = {1\over 16}$.

At $R=\sqrt{2}$, the $S^1/\bZ_2$ theory is equivalent to two copies of the Ising CFT.  At this point, the two twist fields $\sigma_{1,2}$ are the spin operators of the two Ising CFTs. The exactly marginal operator $\cal O$ in \eqref{EMS1Z2} becomes
\ie
{\cal O}(z,\bar z)  =\varepsilon_1(z,\bar z)\, \varepsilon_2(z,\bar z)\,,
\fe
where $\varepsilon_i(z,\bar z)$ is the energy operator of weight $(h, \bar h)=(\frac 12, \frac 12)$ of the $i$-th Ising CFT.

\paragraph{$\bZ_2^{\cal B}$ Orbifold of $S^1/\bZ_2$}

At a generic radius of  the $S^1/\bZ_2$ theory with radius $R\ge 1$, the theory has a $\bZ_2^{\cal B}$ symmetry \cite{Dijkgraaf:1987vp}:\footnote{If we choose to describe the same theory in the T-dual frame with radius $R^{\rm T} = 1/R \le1$, then the $\bZ_2^{\cal B}$ symmetry acts on $V_{n,w}$ in the T-dual frame by a phase $(-1)^w$ because T-duality exchanges $n$ with $w$.}
\ie\label{Z2BS1Z2}
&\bZ_2^{\cal B} :~~\sigma_1 \to -\sigma_1\,,~~~~\sigma_2 \to \sigma_2\,,~~~~V_{n,w} \to (-1)^n \,V_{n,w}\,.
\fe
At the Ising$^2$ point, $\bZ_2^{\cal B}$ is just the $\bZ_2$ symmetry of one of the Ising CFT.   
 The theory enjoys the  Kramers-Wannier duality for each copy of the Ising CFT:
\ie
{S^1/\bZ_2\left[R={\sqrt{2}}\right]  \over  \bZ_2^{\cal B} }=S^1/\bZ_2[R={\sqrt{2}}]  \,.
\fe
The Kramers-Wannier duality is implemented by a duality defect $\cal N$, which flips the sign of $\varepsilon_1$ and maps the order operator $\sigma_1$ to the disorder operator $\mu_1$.  

The exactly marginal deformation ${\cal O}=\varepsilon_1\varepsilon_2$ is odd the duality defect $\cal N$.   
This implies that starting from the ${\cal B}_0\equiv $ Ising$^2$ point, the theory ${\cal B}$ deformed by $+{\cal O}$ and the theory ${\cal B}'$ deformed by $-{\cal O}$ are related to each other by the $\bZ_2^{\cal B}$ orbifold. 
The radii of the theories on two sides can be worked out to be\footnote{ Using T-duality we can rewrite \eqref{KWS1Z2} as ${ {S^1/\bZ_2}[R]\over \bZ_2^{\cal B} }  = S^1/\bZ_2 \left[ R/2\right ]$. However, the definition of $\bZ_2^{\cal B}$ in \eqref{Z2BS1Z2} is not T-duality invariant, so ${ {S^1/\bZ_2}[R]\over \bZ_2^{\cal B} } \neq { {S^1/\bZ_2}[1/R]\over \bZ_2^{\cal B} } $.}
\ie\label{KWS1Z2}
{ {S^1/\bZ_2}[R]\over \bZ_2^{\cal B} }  = S^1/\bZ_2 \left[{2\over R}\right ]\,~~~~(R\ge1)\,.
\fe
We show this equality explicitly at the level of the torus partition function in Appendix \ref{app:ZS1Z2}.

\subsection{Bosonic $T^2$ CFT}\label{sec:T2}

The third example is the $c=2$ CFT whose target space is a torus $T^2$. The conformal manifold is four-dimensional, and we will identify a particular locus along which the family of CFTs are related to each other by $\bZ_2^{\cal B}$ orbifold.  Our exposition follows \cite{Polchinski:1998rq} (see also \cite{Gukov:2002nw} for the classification of rational points on the conformal manifold of the $T^2$ CFT). 

We will normalize the two scalar fields to have periodicities $X^1(z,\bar z) \sim X^1(z,\bar z)+2\pi R\,,X^2 (z,\bar z)\sim X^2(z,\bar z)+2\pi R\,.$ 
The metric and the $B$ field of the $T^2$ CFT will be denoted as $G_{ij}$ and $B_{ij}$ with $i,j=1,2$, parametrizing the conformal manifold of the $T^2$ CFT.   Since we only have two scalars, there is only one $B$ field, $b\equiv B_{12}$.  The $B$ field modulus is periodic, $b\sim b+1$.

The metric moduli includes the K\"ahler modulus $R$ and the complex structure moduli $\uptau$. The latter is encoded in $G_{ij}$ as
\begin{align}
G_{ij} = \left(\begin{array}{cc}1 & \uptau_1 \\\uptau_1 & |\uptau|^2\end{array}\right)\,,
\end{align}
where $\uptau = \uptau_1 + \ii\uptau_2$ and $|\uptau|^2 = \uptau_1^2 +\uptau_2^2$.\footnote{To distinguish the target space torus from the spacetime torus, we use $\uptau$ for the complex structure of the former, while  $\tau$ for the latter. For the spacetime torus, we also use   $q=e^{2\pi \ii \tau}$.}  The complex structure moduli $\uptau$ are subject to the $PSL(2,\bZ)$ identification. 
Let us summarize the exactly marginal deformations of the $T^2$ CFT:
\ie
R>0\,,~~~\uptau \sim {{\rm a}\uptau+{\rm b}\over {\rm c}\uptau+{\rm d}}\,,~~~b\sim b+1\,,
\fe
where ${\rm a,b,c,d}\in \bZ$ and ${\rm ad}-{\rm bc}=1$. There are also T-duality identifications but we will not discuss them here.

\paragraph{Primary Operators}

The local primary operators $V_{n_1,w^1,n_2,w^2}(z,\bar z)$ of the ${\mathfrak{u}(1)}_L^2\times {\mathfrak{u}(1)}_R^2$ current algebra are labeled by four integers, two momentum numbers $n_1,n_2$ and two winding numbers $w^1,w^2$. 
Its conformal weights are given as follows. Let
\begin{align}
v_i \equiv { n_i \over R} - B_{ij} w^j R\,,
\end{align}
   Next we define $v_{L}^i= v^i + w^i R\,,v_R^i  = v^i -w^iR$, 
where the indices are raised and lowered by $G_{ij}$ and $G^{ij}$.
The conformal weights of $V_{n_1,w^1,n_2,w^2}$ are
\begin{align}\label{T2hh}
h  = \frac14 G_{ij} v^i_L v^j_L\,,~~~~~\bar h  = \frac 14 G_{ij} v^i_Rv^j_R\,.
\end{align}
The Lorentz spin of the operator is $s= h-\bar h  =  n_i w^i$. 
The primary operator $V_{n_1,w^1,n_2,w^2}$  can be written in terms of the left- and right-moving compact bosons as
\ie\label{O}
V_{n_1,w^1,n_2,w^2}(z,\bar z) = 
\exp\left[\,
\sum_{i=1}^2  \, \ii v_L^i  X_L^i  (z) +\ii v_R^i  X_R^i  (\bar z) 
\,\right]
\fe

There are  special points on the conformal manifold where the CFT is described by the WZW model: 
\begin{itemize}
\item ${\mathfrak{su}(2)}_1\times{\mathfrak{su}(2)}_1 ={\mathfrak{so}(4)}_1  : ~~R=1,\, (\uptau_1,\uptau_2 )= (0,1) , \,b=0.$
\item ${\mathfrak{su}(3)}_1:~~ R=1, \, ( \uptau_1,\uptau_2 ) = (\frac12, -{\sqrt{3}\over2}), \,b=1/2.$
\item ${\mathfrak{su}(2)}_1\times {\mathfrak{u}(1)}_6 :~~ R=1,\,  (\uptau_1= 0, \uptau_2= \sqrt{3} ),\, b=0. $
\end{itemize}

\paragraph{$\bZ_2^{\cal B}$ Self-Dual Locus}

We will be interested in a particular non-anomalous $\bZ_2^{\cal B}$ global symmetry that exists at any point on the conformal manifold of the $T^2$ CFT. 
Its action on the local operator is  
\ie
\bZ_2^{\cal B} :~V_{n_1,w^1,n_2,w^2 } \to (-1)^{n_1+w^1+n_2+w^2} V_{n_1,w^1,n_2,w^2}\,.
\fe
At the special point  of ${\mathfrak{su}(2)}_1\times{\mathfrak{su}(2)}_1$, this $\bZ_2^{\cal B}$ symmetry is the diagonal subgroup of the center $\bZ_2$'s of the two left-moving ${\mathfrak{su}(2)}$'s.  
We discuss the twisted torus partition functions of the $T^2$ CFT with respect to $\bZ_2^{\cal B}$ in Appendix \ref{app:ZT2}.

Now consider two one-dimensional loci $\cal B$ and ${\cal B}'$ on the conformal manifold, joining at the ${\mathfrak{su}(2)}_1\times {\mathfrak{su}(2)}_1$ point $(R=1 , \uptau_1=0 ,\uptau_2= 1, b=0)$:\footnote{The theory ${\cal B}[\uptau_2]$ with $\uptau_2\le1$ is identical to a theory with $\uptau_2\ge1$ by T-duality, and similarly for ${\cal B}'[b]$ with $b<0$. }
\begin{align}
&{\cal B}[\uptau_2]:~~R=1\,,~~~\uptau_1=0\,,~~~~\uptau_2\ge 1\,,~ ~~~~~~~~~~~~b=0\,,\nonumber\\
&{\cal B}'[b]:~~\,R=1\,, ~~~\uptau_1=b\,,~~~~\, \uptau_2=\sqrt{1-{b}^2}\,,~~~1> b\ge0\,.
\end{align}
The two families of theories $\cal B$  and ${\cal B}'$ are obtained from ${\mathfrak{su}(2)}_1\times {\mathfrak{su}(2)}_1$ by the following two exactly marginal deformations ${\cal O}$ and ${\cal O}'$, respectively:
\ie\label{OOp}
&{\cal O} (z,\bar z)=  \partial X^2(z) \bar\partial X^2 (\bar z)\,,\\
&{\cal O}' (z,\bar z)=  \partial X^2(z) \bar\partial X^1 (\bar z)\,.
\fe
Note that  each  ${\cal O}$ and ${\cal O}'$ preserves a copy of the $ \mathfrak{su}(2) \times \mathfrak{u}(1)\times\overline{ \mathfrak{su}(2) }\times \overline{\mathfrak{u}(1)}$ current algebra, but they preserve different subalgebras of $\mathfrak{su}(2)\times \mathfrak{su}(2) \times \overline{\mathfrak{su}(2)}\times\overline{ \mathfrak{su}(2) }$.  
More explicitly, the  $\mathfrak{su}(2) \times \overline{\mathfrak{su}(2)}$ currents along the ${\cal B}[\uptau_2]$ path are:
\ie
{\cal B}[\uptau_2]:&~~~(h,\bar h) = (1,0):~~~ J_L^{\,\rm x}\pm  \ii J_L^{\,\rm y}\sim  V_{ (\pm 1,\pm 1,0,0)}\,,\\
&~~~~~~~~~~~~~~~~~~~~~~~~~~J_L^{\,\rm z} \sim \ii \partial X^1(z)\,,\\
&~~~(h,\bar h) = (0,1):~~~ J^{\,\rm x}_R\pm  \ii J_R^{\,\rm y}\sim V_{ (\pm 1,\mp 1,0,0)}\,, \\
&~~~~~~~~~~~~~~~~~~~~~~~~~~J_R^{\,\rm z} \sim \ii \bar \partial X^1(\bar z)\,.
\fe
On the other hand, those on the ${\cal B}'[b]$ path are
\ie
{\cal B}'[b]:&~~~(h,\bar h) = (1,0):~~~ J_L^{\,\rm x}\pm  \ii J_L^{\,\rm y}\sim V_{ (\pm 1,\pm 1,0,0)}\,,\\
&~~~~~~~~~~~~~~~~~~~~~~~~~~J_L^{\,\rm z} \sim \ii \partial X^1(z)\,,\\
&~~~(h,\bar h) = (0,1):~~~J^{\,\rm x}_R\pm  \ii J_R^{\,\rm y}\sim V_{  (0,0,\pm1,\mp 1)}\,,\\
&~~~~~~~~~~~~~~~~~~~~~~~~~~J_R^{\,\rm z} \sim \ii\bar \partial X^2(\bar z)\,.
\fe

\begin{figure}
\centering
\includegraphics[width=.5\textwidth]{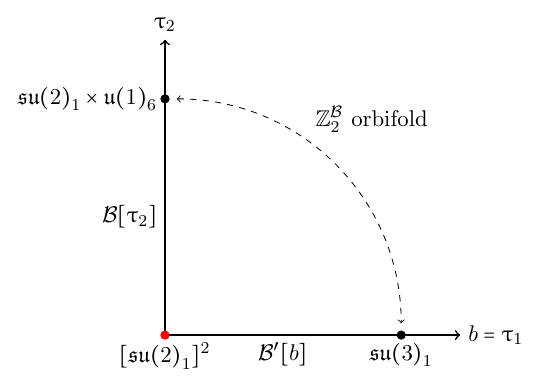}
\caption{A one-dimensional locus on the conformal manifold of the $T^2$ CFT.  The two families of CFTs ${\cal B}[\uptau_2]$ and ${\cal B}'[b]$ are related by gauging $\bZ_2^{\cal B}$  point-by-point with $\uptau_2 = \sqrt{1+b\over 1-b}$.}\label{fig:L}
\end{figure}

Below we will show that 
\ie\label{T2KW}
 {\cal B}'[b]=  {{\cal B}[\uptau_2]\over\bZ_2^{\cal B}  } 
 \,,~~~~\uptau_2 = \sqrt{1+b\over 1-b}\,.
\fe
The structure of this one-dimensional locus on the conformal manifold is shown in Figure \ref{fig:L}. 
To show \eqref{T2KW}, we will write down a one-to-one map between the operators $V'_{n_1',{w'}^1,n'_2,{w'}^2}$ of ${\cal B}'[b]$  and the untwisted and the twisted sectors of the orbifold theory ${\cal B}[\uptau_2] /\bZ_2^{\cal B}$ for all $\uptau_2 \ge 1$.  
Let us start with the untwisted sector, which consists of $\bZ_2^{\cal B}$ even operators $V_{n_1,w^1,n_2,w^2}$ satisfying $n_1+w^1+n_2+w^2\in 2\bZ$.  
The untwisted sector operators are mapped to  $V'_{n_1',{w'}^1,n'_2,{w'}^2}$ of ${\cal B}'[b]$ as
\ie\label{untwmap}
\begin{split}
&n_1'  = \frac 12 (n_1 +w^1 -n_2 +w^2) \,,\\
&{w'}^1 = \frac 12 (n_1+w^1+n_2-w^2)\,,\\
&n_2'  = \frac 12 (n_1 -w^1 -n_2 -w^2) \,,\\
&{w'}^2 = \frac 12 (-n_1+w^1-n_2-w^2)\,.
\end{split}
\fe
Note that since the untwisted sector operators satisfy $n_1+w^1+n_2+w^2\in 2\bZ$, the resulting $n_i',{w'}^i$ are integers.  
Furthermore, $n_1' + {w'}^1 + n_2' +{w'}^2 \in 2\bZ$, so these operators are even under the ${\bZ_2^{\cal B}}'$ symmetry in ${\cal B}'[b]$.

The rest of the $V'_{n_1',{w'}^1,n'_2,{w'}^2}$ operators come from the $\bZ_2^{\cal B}$-even,  twisted sector of ${\cal B}/\bZ_2^{\cal B}$, which consists of operators $V_{\tilde n_1, \tilde w^1 ,\tilde n_2,\tilde w^2}$ with $\tilde n_i,\tilde w^i \in \frac 12+\bZ$ and $\tilde n_1\tilde w^1 +\tilde n_2\tilde w^2 \in \bZ$.  The latter two conditions imply that $\tilde n_1, \tilde w^1 ,\tilde n_2,\tilde w^2 \in 2\bZ+1$.  
The twisted sector operator are mapped to  ${V}'_{n_1',{w'}^1,n'_2,{w'}^2}$ of ${\cal B}'[b]$ by the same map \eqref{untwmap} but with tilde  on the righthand side.  These ${V}'_{n_1',{w'}^1,n'_2,{w'}^2}$'s have $n_1' + {w'}^1 + n_2' +{w'}^2 \in 2\bZ+1$, so they are odd under the ${\bZ_2^{\cal B}}'$ symmetry in ${\cal B}'[b]$.

We have therefore shown that the spectrum of local operators of ${\cal B}[\uptau_2]/\bZ_2^{\cal B}$ is isomorphic to ${\cal B}'[b]$ with the moduli identified as $\uptau_2 = \sqrt{1+b\over 1-b}$.

The duality  between $\cal B$ and ${\cal B}'$ is implemented by a (0+1)-dimensional duality interface.
 At the  ${\mathfrak{su}(2)}_1\times {\mathfrak{su}(2)}_1$ point, this duality interface becomes a duality defect $\cal N$.  
From \eqref{OOp}, we see that ${\cal O}$ and ${\cal O}'$ are related by exchanging the $X_R^1$ with $X_R^2$ (but leaving $X_L^{1,2}$ as they were). 
Naively, one might think this exchange action is a  $\bZ_2$ global symmetry of the full (bosonic) theory. 
This is, however, not true, because we cannot consistently extend such an exchange action to an invertible map from local operators to local operators.  
For example,  this exchange action would have mapped the primary operator ${V}_{n_1,w^1,n_2,w^2}$ to
\ie\label{exchange}
{V}_{ { n_1+w^1 +n_2-w^2\over2} ,  { n_1+w^1 -n_2+w^2\over2}  , { n_2+w^2 +n_1-w^1\over2} ,{ n_2+w^2 -n_1+w^1\over2} }\,.
\fe
Similar to the compact boson CFT, the righthand side is only a local operator if $n_1+w^1+n_2+w^2 \in 2\bZ$, {\it i.e.} when ${V}_{n_1,w^1,n_2,w^2}$ is $\bZ_2^{\cal B}$ even.  
On the other hand, when ${V}_{n_1,w^1,n_2,w^2}$ is $\bZ_2^{\cal B}$ odd, it is mapped to a non-local operator in the twisted sector of $\bZ_2^{\cal B}$.   
Therefore, we conclude that $\cal N$ acts on local operators in a non-invertible way, which is a characteristic feature of a duality defect.

\section{Fermionic Models}\label{sec:Fmodels}

In this section, we discuss the fermionic dual of the Kramers-Wannier transitions on the bosonic conformal manifold. 
See \cite{Karch:2019lnn,Gaiotto:2019gef} for discussions on the $c=1$ fermionic (spin) CFTs. 

\subsection{Dirac Fermion}\label{subsec:Dirac}

The simplest example of  a topological  transition on the conformal manifold is to take ${\cal F}_0$ to be the $c=1$ free massless Dirac fermion, which is equivalent to two left-moving Majorana fermions $\psi_L^i(z)$ and two right-moving Majorana fermions $\psi_R^i(\bar z)$, with $i=1,2$.\footnote{As a fermionic theory, there is no distinction between a Dirac fermion and two Majorana fermions.  However, there are different ways to sum over the spin structures when trying to obtain a bosonic theory.  In  \cite{Elitzur:1986ye},  the authors use  ``Dirac fermion" and  ``two Majorana fermions" to refer to  two different ways of summing over the spin structures, corresponding to the $R=\sqrt{2}$ compact boson theory (discussed in Section \ref{Sec:FreeBoson}) and the $R=\sqrt{2}$  orbifold theory $S^1/\bZ_2$ (discussed in Section \ref{sec:S1Z2}), respectively.}  
 Let $\Psi_{L,R}\equiv \psi_{L,R}^1 + \ii \psi_{L,R}^2$, $\Psi_{L,R}^\dagger\equiv \psi_{L,R}^1 - \ii \psi_{L,R}^2 $. 
   The theory has a left and a right $\mathfrak{u}(1)\times \overline{\mathfrak{u}(1)}$ current algebras generated by $\Psi_L(z)\Psi_L^\dagger(z)$ and $\Psi_R(\bar z)\Psi_R^\dagger(\bar z)$, respectively. 
The Dirac fermion theory is the fermionization of the $c=1$ compact boson discussed in Section \ref{Sec:FreeBoson} at $R=\sqrt{2}$ with respect to the $\bZ_2^{\cal B}=\bZ_2^{(1,0)}$ symmetry defined in \eqref{Z2BS1}.

There is one exactly marginal deformation, the Thirring deformation:
\ie\label{Thirring}
{\cal O}(z,\bar z) = \Psi_L(z) \Psi_L^\dagger(z)  \Psi_R(\bar z) \Psi_R^\dagger (\bar z)\,.
\fe
The relation between the Thirring coupling and the radius $R$ of the compact boson theory was derived in \cite{Coleman:1974bu}. 
We will therefore use $R$ to denote exactly marginal coupling for \eqref{Thirring} and denote the deformed Dirac fermion theory as ${\rm Dirac}[R]$. 
In particular, ${\rm Dirac}[\sqrt{2}]$ is the free, massless Dirac fermion.

\begin{figure*}
\centering
\includegraphics[width=.6\textwidth]{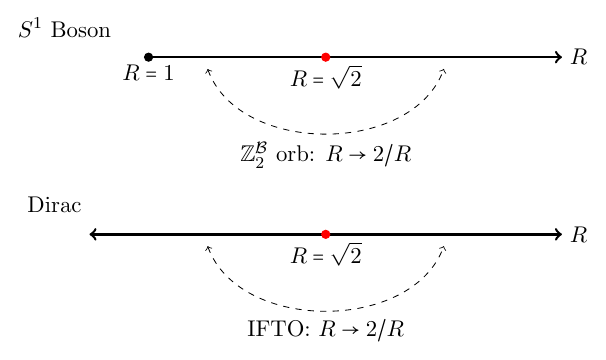}
\caption{The conformal manifold of the  $c=1$ compact boson theory $S^1[R]$ (top) and that of the Dirac fermion perturbed by the Thirring coupling  Dirac$[R]$ (bottom). The former is a half-line, while the latter is a full line.}\label{fig:dirac}
\end{figure*}

\paragraph{Topology of the Dirac Branch}
Let us comment on the global topology of the conformal manifold for the Dirac fermion branch parametrized by $R$.  
We start by noting that, due to the T-duality $S^1[R]=S^1[ 1/R]$,  the $S^1$ branch of the compact boson theory is  a half-line, with the endpoint located at $R=1$, {\it i.e.} the $\mathfrak{su}(2)_1$ WZW model.  
On the fermion side, ${\rm Dirac}[R]$ is the fermionization of $S^1[R]$ with respect to  $\bZ_2^{\cal B} =\bZ_2^{(1,0)}$.   
Since the momentum $\bZ_2^{\cal B}=\bZ_2^{(1,0)}$  is exchanged with the winding $\bZ_2^{(0,1)}$  under the T-duality, the fermionization does \textit{not} commutes with the T-duality of the bosonic theories.  Therefore ${\rm Dirac}[R]  \neq {\rm Dirac}[1/R]$.\footnote{Put differently, we can define another family of fermionic theories, denoted as $\widetilde{\rm Dirac}[R]$, by fermionizing  $S^1[R]$ with respect to $\bZ_2^{(0,1)}$. Then $\widetilde{\rm Dirac}[R] = {\rm Dirac}[1/R]$.}  
Consequently, the topology of the Dirac branch of the $c=1$ fermionic CFT is  $\bR$ instead of  a half line  \cite{Karch:2019lnn}.\footnote{One can alternatively regard the IFTO as a (1+1)$d$ local counterterm, and identify  Dirac[$R$] with Dirac[$2/R$].  From this perspective the Dirac branch of the fermionic conformal manifold is again a half line, but the origin is now located at the Dirac point $R=\sqrt{2}$ instead of the $\mathfrak{su}(2)_1$ point $R=1$. } 
See Figure \ref{fig:dirac}.

\paragraph{$\bZ_2^{\rm IFTO}$ Symmetry}
The $\bZ_2^{\rm IFTO}$ symmetry is defined as
\ie\label{Z2IFTO1}
\bZ_2^{\rm IFTO}:~~&\Psi_L(z) \to \Psi_L^\dagger(z)\,,~~~~~\Psi_L^\dagger(z) \to \Psi_L(z)\,,\\
&\Psi_R(\bar z) \to \Psi_R(\bar z)\,,~~~~~\Psi_R^\dagger(\bar z)\to \Psi_R^\dagger(\bar z)\,,
\fe
which is the particle-hole transformation  $\bZ_2^{C_L}$  on the left-moving fermion operators. 
Given  a fixed spin structure, if we treat the Dirac fermion as two Majorana fermions, then $\bZ_2^{\rm IFTO}$ is the chiral fermion parity $(-1)^{F_L}$ for one copy of the Majorana fermion.  
Note that the exactly marginal operator $\cal O$ is $\bZ_2^{\rm IFTO}$-odd.

In Section \ref{Sec:FreeBoson} we showed that the bosonic theory $S^1[R]$ is related to $S^1[2/R]$ by the $\bZ_2^{\cal B}$ orbifold.  
It follows from the commutative diagram in Figure \ref{fig:CD} that:
\ie
Z_{\rm Dirac}[\rho,R] =  Z_{\rm IFTO} [\rho]  Z_{\rm Dirac}[\rho, 2/R]\,,
\fe
 Hence as we move along the one-dimensional conformal manifold generated by $\cal O$ form $R<\sqrt{2}$ to $R>\sqrt{2}$, the models on the two sides $\cal F$ and ${\cal F}'$ of a free massless Dirac fermion differ by an IFTO \cite{Karch:2019lnn}. 
This is our first example of a topological transition on the fermionic conformal manifold. 
 Table \ref{VnwF} shows the operator spectrum of $c=1$ fermionic CFTs.

Let us describe the local operators of the fermionic model in terms of $V_{n,w}$. 
In the anti-periodic $(A)$ sector of the Dirac fermion, the local operators come from (1) the $\bZ_2^{\cal B}$-even, local operators plus (2) the $\bZ_2^{\cal B}$-odd non-local operators from the $\bZ_2^{\cal B}$ twisted sector (see \eqref{torusBF}).  The former are operators $V_{n,w}$ with $n\in 2\bZ$, $w\in \bZ$,  while the latter are $V_{\tilde n, \tilde w}$ with $\tilde n \in 2\bZ+1 $, $\tilde w\in \frac 12+\bZ$.  
We have used the fact that the twisted sector operator $V_{\tilde n,\tilde w}$ is $\bZ_2^{\cal B}$-even  if the spin $s=\tilde n \tilde w$ is an integer, while it is $\bZ_2^{\cal B}$-odd if $s$ is a half-integer \cite{Hung:2013cda,Chang:2018iay,Lin:2019kpn}. 
We  identify the  fermion operators as
\begin{align}
\begin{split}
&\Psi_L (z)= V_{ 1, \frac{1}{2}}\,, ~~ \Psi^\dagger_L (z)=V_{-1,-\frac{1}{2}}\,,~~(h,\bar h)=\left(\frac{1}{2},0\right)\,,\\
&\Psi_R (\bar z)= V_{ 1, -\frac{1}{2}}\,, ~~ \Psi^\dagger_R (\bar z)=V_{-1,\frac{1}{2}}\,,~~(h,\bar h)=\left(0,\frac{1}{2}\right)\,.
\end{split}
\end{align}

It is then straightforward to see that $\bZ_2^{\rm IFTO}$ acts as
\ie\label{Z2NDirac}
\bZ_2^{\rm IFTO}:~~V_{n,w} (z,\bar z) \to V_{ -2w , - \frac n2}(z,\bar z)\,.
\fe
For example, $\bZ_2^{\rm IFTO}$ exchanges $V_{1,1/2}=\Psi_L(z)$ with $V_{-1,-1/2}=\Psi_L^\dagger(z)$, but leaves $V_{1,-1/2}=\Psi_R(\bar z)$ with $V_{-1,1/2}=\Psi_R^\dagger(\bar z)$ invariant.  
Note that $\bZ_2^{\rm IFTO}$ anticommutes with the left-moving current algebra generator $\Psi_L(z)\Psi^\dagger_L(z)$ but  commutes with the right-moving one $\Psi_R(\bar z)\Psi^\dagger_R(\bar z)$.

What happens when we extend the $\bZ_2^{\rm IFTO}$ symmetry in the fermionic theory to its bosonization? 
In the latter, the local primary operators are labeled by {\it integral} $n$ and $w$. 
Hence the $\bZ_2^{\rm IFTO}$ symmetry in the fermionic theory does not extend to an action that maps a local operator to another local operator. 
In fact, it maps the $\bZ_2^{\cal B}$-odd operators ({\it i.e.} those with odd $n$) to a non-local operator in the $\bZ_2^{\cal B}$-twisted sector. 
This is indeed how the non-invertible duality defect $\cal N$ acts as discussed in Section \ref{Sec:FreeBoson}. 
We  have therefore demonstrated how a symmetry action in a fermionic theory is extended to a non-invertible defect under bosonization (see Section \ref{app:ext}). 

\begin{table*}[ht]
\renewcommand{\arraystretch}{1.4}
\centering
\begin{tabular}{ c  c  c  c  c }
\hline
\hline
bosonic sector & fermionic sector & range of $n$ & range of $w$ & example operators\\
\hline
\hline
${\cal H}^E$ & ${\cal H}_A^E$ & $2\ZZ$ & $\ZZ$ & $V_{2,0}=\Psi_L\Psi_R$, $V_{0,-1}=\Psi_L^\dagger \Psi_R$\\
${\cal H}^O$ & ${\cal H}_P^E$ & $2\ZZ+1$ & $\ZZ$ & $O_{\rm ord} = V_{1,0}$\\
$\widetilde{\cal H}^E$ & ${\cal H}_P^O$ & $2\ZZ$ & $\frac{1}{2}+ \ZZ$ & $O_{\rm dis} = V_{0,-\fh}$\\
$\widetilde{\cal H}^O$ & ${\cal H}_A^O$ & $2\ZZ+1$ & $\frac{1}{2}+\ZZ$ & $\Psi_L,\Psi_L^\dagger, \Psi_R, \Psi_R^\dagger$ \\
\hline
\hline
\end{tabular}
\caption{Operators $V_{n,w}$ in the $c=1$ fermionic CFT Dirac$[R]$. This is analogous to \eqref{Isingtable} for the Ising CFT and the Majorana fermion. Together with chiral fermion operator $\Psi_L$ and by operator product expansion, they generate other sectors from ${\cal H}_A^E$, $O_e: {\cal H}_A^E \rightarrow {\cal H}_P^E, O_m: {\cal H}_A^E \rightarrow {\cal H}_P^O, \Psi_L: {\cal H}_A^E\rightarrow {\cal H}_A^O$. 
}\label{VnwF}
\end{table*}

\paragraph{Order and Disorder Operators} 
From the bosonic $S^1[R]$ point of view, it is natural to identify the lightest $\bZ_2^{\cal B}$-odd, local operator  $V_{1,0}$ in ${\cal H}^O$ as the order operator, while the lightest $\bZ_2^{\cal B}$-even, non-local operator $V_{0,-1/2}$ in $\widetilde{\cal H}^E$ as the disorder operator:
\ie
&\text{Order:} ~~~ O_{\rm ord}=V_{1,0}\,,~~~~~~~~~~\Delta_{\rm ord}=  {1\over 2R^2} \\
&\text{Disorder:} ~~~O_{\rm dis}=V_{0,-\fh}\,,~~~~~\Delta_{\rm dis}= {R^2\over8}\,.
\fe
That the order and disorder fields are mutually non-local in the field theory with $U(1)$ symmetry has been studied for long \cite{Marino:1980rn,Koberle:1983gs,marino1983quantum,marino1984phases} in one dimension, and more recently in higher dimension \cite{Marino:1990yi,Marino:2017aio}.
Here we identify the order field as through the study of (non-anomalous) $Z_2$ global symmetry. It does not require $U(1)$ symmetries. 
Note that both  $O_{\rm ord}$ and $O_{\rm dis}$ are scalar operators, {\it i.e.} $s=h-\bar h=0$. 
From the fermionic CFT Dirac$[R]$ point of view, both $O_{\rm ord}$ and $O_{\rm dis}$ are in the $P$ sector, with opposite $(-1)^F$ quantum numbers. 
At the Dirac point $R=\sqrt{2}$, $\bZ_2^{\rm IFTO}$ is a global symmetry that exchanges the order with the disorder operators:
\begin{align}\label{orderdis}
\bZ_2^{\rm IFTO}: ~~ O_{\rm ord}\mapsto O_{\rm dis}\,.
\end{align}
Indeed, the $\bZ_2^{\rm IFTO}$ becomes the Kramers-Wannier duality defect under bosonization.

In the topological transition from Dirac$[R<\sqrt{2}]$ to Dirac[$R>\sqrt{2}]$, both the order and the disorder operators have power law two-point functions  at all radii:
\ie
&\langle O_{\rm ord}(z,\bar z)  O_{\rm ord}(0)\rangle = {1\over |z|^{2\Delta_{\rm ord}}}\,,\\
&\langle O_{\rm dis}(z,\bar z)  O_{\rm dis}(0)\rangle = {1\over |z|^{2\Delta_{\rm dis}}}\,.
\fe
The exponents of the power law fall-off obey
\ie
\Delta_{\rm ord}  >\Delta_{\rm dis}\,, ~~~~~R<\sqrt{2}\,,\\
\Delta_{\rm ord}  < \Delta_{\rm dis}\,, ~~~~~R>\sqrt{2}\,.
\fe
When $R<\sqrt{2}$,   the two-point function of the order operator $O_{\rm ord}$ approaches zero asymptotically faster than that of the disorder operator, and vice versa for $R>\sqrt{2}$. 
This is to be contrasted with the standard second order phase transition, where in one phase the order two-point function falls off exponentially while the disorder two-point function approaches a constant at large distance, and vice versa in the other phase.  See Figure \ref{fig:toptrans}.

\paragraph{Ultraviolet Realization on the Lattice} The complex fermion CFT arises as a the infrared (IR) description of the Luttinger liquid \cite{luther1974backward,theumann1967single,Haldane:1981zza,voit1995one}, a (1+1)$d$ spinless electron system appearing in condensed matters. In the Luttinger liquid at a generic filling, the UV symmetry  is $G_{UV}=U(1)_C\times U(1)_{\text{trn}}$, where $U(1)_C$ is the total charge conservation, $U(1)_{\text{trn}}$ is the translation symmetry.  The operator $V_{n,w}(z,\bar z)$ has charge $q_C=n, q_{\text{trn}}=-2w k_F$, where $k_F$ is the Fermi momentum. This is determined from the fact that $\Psi_L(z)$ carries $q_C=1, q_{\text{trn}}=-k_F$, and $\Psi_R(\bar z)$ carries $q_C=1, q_{\text{trn}}=k_F$. In particular, the fermion bilinear $\Psi_L (z)\Psi_R (\bar z)$ has charge $q_C= 2$ and $\Psi^\dagger_L (z)\Psi_R (\bar z)$ has charge $q_{\text{trn}}= 2k_F$. 
 There is no symmetric relevant operator in the $A$ sector satisfying $\Delta < 2, q_C=q_{\text{trn}}=0$. The Thirring deformation $\calO$ in (\ref{Thirring}) is the symmetric perturbation with the lowest weight. Tuning its coupling from negative to positive through the free Dirac fermion, drives a topological phase transition differing by an IFTO. 

In the lower energy theory of the Luttinger liquid, $V_{2,0}=\Psi_L\Psi_R$ carries $q_C=2, q_{\text{trn}}=0$, representing the superconducting (SC) order operator, while $V_{0,-1} = \Psi_L^\dagger \Psi_R$ carries $q_C=0, q_{\text{trn}}=2k_F$, representing the charge density wave (CDW) order operator.  
At the Dirac point $R=\sqrt{2}$, the $\bZ_2^{\rm IFTO}$ acts as
\begin{align}
\bZ_2^{\rm IFTO}: ~~ V_{2,0}= \Psi_L\Psi_R \, \mapsto\,  V_{0,-1}=\Psi_L^\dagger \Psi_R
\end{align}
This is the well-known duality between the SC order and the CDW order in the Luttinger liquid.  
The duality exchanges  the Luttinger parameter  $K \leftrightarrow 1/K$, which maps the $K<1$ attractive interacting region to the $K>1$ the repulsive interacting region.  
(As we reviewed  in Appendix \ref{app:TL}, the Luttinger parameter $K$ is related to the compact boson radius $R$ as $R=\sqrt{2}K$.)
Meanwhile, the $\bZ_2^{\rm IFTO}$ exchanges the order and the disorder operators as in \eqref{orderdis}. 
We have therefore demonstrated that the attractive/repulsive duality in the bosonized Luttinger liquid is parallel to the Kramers-Wannier duality in the Ising CFT.

\subsection{Majorana $\times$ Ising}\label{sec:MIsing}

Next we consider a fermionic $c=1$ CFT that is the direct product of a single Majorana fermion $\psi_L(z), \psi_R(\bar z)$ and the bosonic Ising CFT, {\it i.e.} ${\cal F}_0 = {\rm Maj}\times {\rm Ising}$.   

The $\bZ_2^{\rm IFTO}$ symmetry is taken to be the left-moving fermion parity $(-1)^{F_L}$ that flips the sign of $\psi_L(z)$ but not that of $\psi_R(\bar z)$:
\ie\label{Z2IFTO2}
\bZ_2^{\rm IFTO}=(-1)^{F_L}:~ \psi_L(z) \to - \psi_L(z)\,,~~~~\psi_R(\bar z)  \to \psi_R(\bar z)\,. 
\fe
The theory ${\cal F}_0$ has one exactly marginal operator
\ie
{\cal O} (z,\bar z) = \psi_L(z)\psi_R(\bar z) \varepsilon(z,\bar z)\,,
\fe
which is odd under $\bZ_2^{\rm IFTO}$. We immediately learn that
\ie
Z_{\text{Maj$\times$Ising}}[\rho,R] =  Z_{\rm IFTO} [\rho]  Z_{\text{Maj$\times$Ising}}[\rho, 2/R]\,,
\fe
where we have used the radius $R$ of $S^1/\bZ_2$ to represent the coupling of $\cal O$. 
Again, we have shown in Section \ref{sec:S1Z2}  that the corresponding bosonic theories $\cal B$ and ${\cal B}'$ are related by the $\bZ_2^{\cal B}$ orbifold.  

There is no continuous internal global symmetry in ${\cal F}_0$. The discrete symmetry is $\mathbb{D}_8$ \cite{Dijkgraaf:1987vp}, whose action on the twist fields is generated by $\sigma_1\to -\sigma_1, \sigma_2\to \sigma_2$ and $\sigma_1\leftrightarrow\sigma_2$.  There is, however, a $\mathbb{D}_8$-invariant relevant deformation, $\varepsilon_1+\varepsilon_2$.

\subsection{Four Majorana Fermions}\label{sec:4maj}

Let us take ${\cal F}_0$ to be  a free theory of four Majorana fermions, which is the fermionization of the $c=2$ model in Section \ref{sec:T2}. 
We will denote the four left-moving (right-moving) Weyl fermions as $\Psi_{L,\pm\frac 12},\Psi_{L,\pm\frac 12}^\dagger$ ($\Psi_{R,\pm\frac 12},\Psi_{R,\pm\frac 12}^\dagger$). (The meaning of the subscripts $\pm\frac12$ will be explained momentarily from the lattice realization.)  
These fermion operators can be written in terms of the exponential operators $V_{n_1,w^1,n_2,w^2}$ in \eqref{O} as 

\ie
&\Psi_{L,\pm \fh}= V_{\pm\fh,\pm \fh, \fh,\fh}=e^{\ii \left(\pm X_L^1+X_L^2\right)} \,,\\
& \Psi_{L,\pm \fh}^\dagger =V_{\pm\fh,\pm \fh, -\fh,-\fh}=e^{\ii \left( \mp X_L^1-X_L^2\right)}\,,\\
 &\Psi_{R,\pm \fh} = V_{\pm\fh,\mp \fh, \fh,-\fh}=e^{\ii \left(\pm X_R^1+X_R^2\right)}\,,\\
& \Psi_{R,\pm \fh}^\dagger =V_{\mp\fh,\pm \fh, -\fh,\fh}=e^{\ii\left(\mp X_R^1-X_R^2\right)}
\fe

The theory ${\cal F}_0$ has a ${\mathfrak{so}(4)}\times\overline{\mathfrak{so}(4)}={\mathfrak{su}(2)}\times {\mathfrak{su}(2)}\times \overline{\mathfrak{su}(2)}\times\overline {\mathfrak{su}(2)}$ current algebra at level 1.  
The $\bZ_2^{\rm IFTO}$ is given in by the duality defect action  \eqref{exchange} in the bosonic model, which exchanges the currents  $\overline{\mathfrak{su}(2)_1}$ with $\overline{\mathfrak{su}(2)_2}$. 
It acts on the fermion by
\ie\label{Z2IFTO3}
\bZ_2^{\rm IFTO}:~~\Psi_{R,-\frac 12 } \leftrightarrow \Psi^\dagger_{R,-\frac 12}
\fe
 leaving the other fermions $\Psi$ invariant.

\begin{figure}
\centering
\includegraphics[width=.4\textwidth]{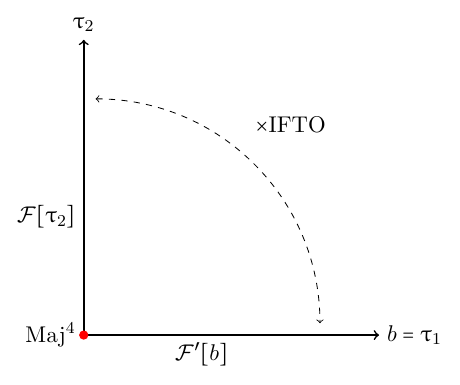}
\caption{A one-dimensional locus on the conformal manifold of four Majorana fermions.  The two families of CFTs ${\cal F}[\uptau_2]$ and ${\cal F}'[b]$ differ by an invertible fermionic topological order.}\label{fig:FL}
\end{figure}

The theory $\calF_0$ has two exactly marginal operators
\begin{align}
\begin{split}
\calO (z,\bar z) \sim& (\Psi_{L,\frac12} \Psi_{L, -\frac 12}^\dagger  +\Psi_{L,-\frac12} \Psi^\dagger_{L,\frac12})\\
&\cdot (\Psi_{R,\frac12} \Psi_{R, \frac 12}^\dagger  +\Psi_{R,-\frac12} \Psi^\dagger_{R,-\frac12})\,,\\
 \calO' (z,\bar z)\sim &(\Psi_{L,\frac12} \Psi_{L, -\frac 12}^\dagger  +\Psi_{L,-\frac12} \Psi^\dagger_{L,\frac12}) \\
&\cdot (\Psi_{R,\frac12} \Psi_{R, \frac 12}^\dagger  -\Psi_{R,-\frac12} \Psi^\dagger_{R,-\frac12})\,,
\end{split}
\end{align}
that are mapped to each other under $\bZ_2^{\rm IFTO}$.  
Note that ${\cal O}$ and ${\cal O}'$ are not proportional to each other  as in the previous examples, but {\it different} operators.
 $\calO$ and $\calO'$ drive the transitions from $\calF_0$ to $\calF$ and $\calF'$, respectively. 
 Following the  discussions  in Section \ref{sec:phase} and in Section \ref{sec:T2}, we conclude that there is a topological transition from ${\cal F}'[b] \to {\cal F}_0\to {\cal F}[\uptau_2]$.  
 In particular, the bosonic relation \eqref{T2KW} is translated into
 \ie
 Z_{{\cal F}'}[  \rho  ,b  ] = Z_{\rm IFTO}[\rho ]  Z_{{\cal F}  }  [ \rho,  \uptau_2 ]\,,~~~~\uptau_2 = \sqrt{1+b\over1-b}\,.
 \fe
 That is, the two families of theories $\cal F$ and ${\cal F}'$ differ by an IFTO point-by-point.

\paragraph{Ultraviolet Realization on the Lattice}

The fermionic CFT can  arise as the infrared  description of a lattice theory in the ultraviolet. 
Let us suppose the UV symmetry on the lattice is $G_{UV}\equiv U(1)_C \times  U(1)_{\text{trn}} \times SU(2)$, where $U(1)_C$ is the total charge conservation, $SU(2)$ is the spin rotation symmetry in a spin-$\fh$ fermion lattice model, and $U(1)_{\text{trn}}$ is the translation symmetry for fermions at incommensurate filling.   The UV symmetry is embedded in the emergent IR symmetries for the theory $\calF$ and $\calF'$ in two different ways. The symmetry generators, actions and the $U(1)_C, U(1)_{\text{trn}}, SU(2)_L, SU(2)_R$ quantum numbers for $V_{n_1,w^1,n_2,w^2}$ are as follows. 
\begin{itemize}
\item $\calF[\uptau_2]$
\begin{align}
\begin{aligned}
&~~J_C=\ii (\partial X^2+\bar\partial X^2),\\
&~~U(1)_C: ~V_{n_1,w^1,n_2,w^2} \rightarrow e^{\ii 2n_2\gamma_C} V_{n_1,w^1,n_2,w^2}\\
&~~J_{\text{trn}}=\ii(\partial X^2-\bar\partial X^2),\\
&~~U(1)_{\text{trn}}:~ V_{n_1,w^1,n_2,w^2} \rightarrow e^{\ii 2w^2\gamma_{\text{trn}}} V_{n_1,w^1,n_2,w^2}\,.\\
\end{aligned}
\end{align}
\begin{align}
&q_C= 2n_2\,,~~~q_{\text{trn}}=2w^2\,,\nonumber\\
& j_L^z=\frac{1}{2}(n_1+w^1)\,,~~~ j_R^z=\frac{1}{2}(n_1-w^1)\,.
\end{align}
\item $\calF'[b]$
\begin{align}
\begin{aligned}
&~~J_C=\ii (\partial X^2+\bar\partial X^1),\\
&~~  U(1)_C: ~V_{n_1,w^1,n_2,w^2} \rightarrow e^{\ii (n_1-w^1+n_2+w^2)\gamma_C} V_{n_1,w^1,n_2,w^2}\\
&~~J_{\text{trn}}=\ii (\partial X^2-\bar\partial X^1),\\
&~~ U(1)_{\text{trn}}:~ V_{n_1,w^1,n_2,w^2}\\
&~~~~~~~~~~~~~~~ \rightarrow e^{\ii (-n_1+w^1+n_2+w^2)\gamma_{\text{trn}}} V_{n_1,w^1,n_2,w^2}\,.
\end{aligned}
\end{align}
\begin{align}
q_C= n_1-w^1+n_2+w^2\,,&~~q_{\text{trn}}=-n_1+w^1+n_2+w^2\,,\notag\\
j_L^z=\frac{1}{2}(n_1+w^1)\,,&~~ j_R^z=\frac{1}{2}(n_2-w^2)\,.
\end{align}
\end{itemize}

\paragraph{Absence of Symmetry-Preserving Relevant Operators} To ensure the perturbative stability of the fermionic theories, we need to exclude  $G_{UV}$-invariant relevant operators  in the anti-periodic ($A$) sector. 
Recall that the theory $\cal F$ and ${\cal F}'$ share the same $A$-sector, {\it i.e.} $Z_\calF[AA]=Z_{\calF'}[AA]$ and  $Z_\calF[AP]=Z_{\calF'}[AP]$.  
From (\ref{torusBF}) and Section \ref{sec:T2}, we learn that the $A$-sector primary operators of $\cal F$ are of the form $V_{n_1,w^1,n_2,w^2}$ satisfying
\begin{align}
&(i) ~ n_i,w^i\in\ZZ,~~ n_1+w^1+n_2+w^2\in 2\ZZ,\nn\\
&(ii) ~ n_i,w^i\in \frac{1}{2}+\ZZ,~~ n_1w^1+n_2w^2\in 1+2\ZZ
\end{align}
The $G_{UV}$-invariant operators further satisfy $q_C=0, q_{\text{trn}}=0, j_L^z+j_R^z=0$, which implies
\begin{align}
\calF:~~ &n_1=n_2=w^2=0\,,\nn\\
\calF':~~ &n_1=w^1=-n_2=w^2\,.
\end{align}
The lightest ({\it i.e.} the smallest scaling dimension)  scalar operators satisfying the above constraints are
\begin{align}
\calF:~~ & V_{0,2,0,0}\,,~~~(h, \bar h)=(1,1)\,,\nn\\
\calF':~~ & V_{1,1,-1,1}\,,~~~ (h, \bar h)=(1,1)\,,
\end{align}
which are marginal but not relevant. 
Therefore, we conclude that in either $\calF$ or $\calF'$, there is no relevant operator neutral under the microscopic symmetry $G_{UV}$. Both theories describe (1+1)$d$ symmetry-protected gapless phases.

\paragraph{Topological Transition and Symmetry Embedding} 
A given embedding of the UV symmetry $G_{UV}$ into the IR emergent symmetry is consistent with either the fermionic CFT $\calF$ or $\calF'$, but not both. Hence a fermionic lattice model can realize either $\calF$ or $\calF'$, but \emph{not} the topological transition from $\calF$ to $\calF'$. However, if there is no symmetry constraint, both theories can describe the multi-critical theories of the same lattice model. 
In this case, by tuning the exactly marginal perturbations $\calO, \calO'$ along the path described by Figure \ref{fig:FL}, one can realize the topological transition from the lattice model.

\section{Duality Defect and the Chiral Fermion Parity}\label{app:ext}

The duality defect $\cal N$ arises from the $\bZ_2^{\rm IFTO}$ line in ${\cal F}_0$ before gauging $(-1)^F$. 
Due to the mixed anomaly between $\bZ_2^{\rm IFTO}$ and $(-1)^F$, the former becomes a non-invertible topological defect $\cal N$ in the gauged theory ${\cal B}_0$ \cite{Bhardwaj:2016clt,Thorngren:2018bhj}. 
This is a generalization to the symmetry extension in gauging a bosonic global symmetry in (1+1) dimensions with mixed anomaly, which we will briefly discuss below.\footnote{We thank Kantaro Ohmori and Yuji Tachikawa for discussions on this point.} 

We start with an example of the usual symmetry extension from the mixed anomaly. 
Consider the $c=1$ compact boson CFT and let   $\bZ_2^{(1,0)}$ and $\bZ_2^{(0,1)}$ be the $\pi$ rotations of the momentum and winding symmetries, respectively. While $\bZ_2^{(1,0)}$ and $\bZ_2^{(0,1)}$ by themselves are free of anomalies, there is a mixed anomaly between the two. 
Now suppose we gauge $\bZ_2^{(1,0)}$, then we throw away operators with odd momentum number $n\in 2\bZ+1$, but introduce operators with half-integral winding numbers $w\in \frac 12+\bZ$ from the twisted sector. It follows that the winding $\bZ_2^{(0,1)}$ acts by a factor of $\pm \ii$ on these new twisted sector states, and hence it is extended to $\bZ_4$ in the gauged theory.

In the case of gauging the $(-1)^F$ of ${\cal F}_0$, we might introduce twisted sector states that might not have a well-defined $\bZ_2^{\rm IFTO}$ action. 
In the example of  a single massless Majorana fermion, the $\bZ_2^{\rm IFTO}$ symmetry is the chiral fermion parity $ (-1)^{F_L}$. 
In the corresponding bosonic theory, {\it i.e.} the Ising CFT, the would-be $\bZ_2^{\rm IFTO}$ acts by a sign on the energy operator $\varepsilon(z,\bar z)= \psi_L(z)\psi_R(\bar z)$. 
However, this is incompatible with the fusion rule of the primary operators, $\sigma\times \sigma = 1+\varepsilon$.  
The obstruction is precisely that there is no invertible action of the would-be $\bZ_2^{\rm IFTO}$ on the spin operator $\sigma$, which comes from the twisted sector when gauging ${\cal F}_0$.  
Hence there is no extension of $\bZ_2^{\rm IFTO}$ in ${\cal F}_0$ to  any bigger group that can be consistent with the above fusion rule. 
Instead,  the  $\bZ_2^{\rm IFTO}$ of ${\cal F}_0$ is extended to a non-invertible defect $\cal N$ of ${\cal B}_0$, satisfying a non-group-like fusion rules \eqref{TYZ2} (see Figure \ref{fig:sym}). 
One can show that the duality defect $\cal N$ is compatible with the fusion rule of local operators $\sigma\times\sigma=1+\varepsilon$ \cite{Frohlich:2004ef,Chang:2018iay}.

More generally, consider $N$ Majorana fermions.  Let $(-1)^{F_L}$ be the chiral
fermion parity  that flips the signs of all the left-moving fermions.
The (1+1)$d$ fermionic 't Hooft anomaly of  $(-1)^{F_L}$ is $N$ mod 8 \cite{PhysRevB.85.245132,Qi_2013,PhysRevB.88.064507,PhysRevB.89.201113,Kapustin:2014dxa,FH160406527}.
 The
bosonization is the bosonic $Spin(N)_1$ WZW model, in which  the
chiral fermion parity $(-1)^{F_L}$ has become
\cite{Numasawa:2017crf,Lin:2019kpn}:\footnote{The fusion category of the
$Spin(N)_1$ WZW model can also be read off from the  modular tensor category for the $Spin(N)_1$ Chern-Simons theory (which is a non-spin TQFT)  by forgetting the braiding. The latter can be found, for example, in Table 1-3 of
\cite{K062}. Note that the periodicity of $N$ for the former is 8, while that for the latter is 16.}
\begin{itemize}
\item $N=1,7$ mod 8: a duality line $\cal N$ obeying the $\bZ_2$ Tambara-Yamagami category TY$_+$ \cite{TAMBARA1998692}. 
\item $N=3,5$ mod 8: a duality line $\cal N$ obeying the other $\bZ_2$ Tamabara-Yamagami category TY$_-$. 
\item $N=0$ mod 8: a $\bZ_2$ line that  is non-anomalous.
\item $N=4$ mod 8:  a $\bZ_2$ line that   is anomalous, corresponding to the nontrivial element of $H^3(\bZ_2,U(1))=\bZ_2$. 
\item $N=2,6$ mod 8: a $\bZ_4$ line. The $\bZ_4$ is anomalous, corresponding to the square of the generator of $H^3(\bZ_4,U(1))=\bZ_4$. 
\end{itemize} 
Both TY$_\pm$ share the same fusion rules \eqref{TYZ2}, but different $F$ symbols.  TY$_+$ is realized by the Ising CFT, while TY$_-$ is realized by the $\mathfrak{su}(2)_2=Spin(3)_1$ WZW model.  

Note that in our discussion of the three fermionic models in Section \ref{sec:Fmodels}, our choice of the $\bZ_2^{\rm IFTO}$ (see \eqref{Z2IFTO1}, \eqref{Z2IFTO2}, \eqref{Z2IFTO3}) always flips the sign of a {\it single}  Majorana-Weyl fermion. Hence our $\bZ_2^{\rm IFTO}$ has 't Hooft anomaly 1 mod 8 and turns into a duality defect TY$_+$ as in the $N=1$ mod 8 case above.

\section{Outlook}\label{sec:conclu}

A natural configuration for the topological transition is a (1+1)$d$ system
with  a (0+1)$d$ interface separating a fermionic CFT $\cal F$ with ${\cal F}'
= {\cal F} \otimes {\rm IFTO}$ as in Figure \ref{fig:dualityinterface}.  
Since the two fermionic CFTs $\cal F$ and ${\cal F}'$ differ by an  IFTO, one
may wonder if the domain wall between $\cal F$ and ${\cal F}'$ contains a
Majorana zero mode or not.  This question can be approached by DMRG
\cite{He:2018hnt}, tensor network \cite{Bal:2018wbw}, MERA \cite{Bridgeman:2017etx} or by the conformal interface \cite{fuchs2007topological}.  

We can also study a (1+1)$d$ gas of spinless fermions on an open chain
with attractive or repulsive interactions.
We know that a single IFTO on an open chain has a two-fold topological degeneracy,
which comes from two Majorana zero modes from the two ends of
the chain.  Here the topological degeneracy has an energy splitting of order
$e^{-L/\xi}$, where $L$ is the length of the chain and $\xi$ is a length scale.  
  Since the fermion system with
attractive interaction and that with repulsive interaction differ
by an IFTO, one may wonder if one of the above two systems might have
topological degeneracy ({\it i.e.} the Majorana zero modes) localized at the chain
ends. More generally, an energy splitting of order
$O\left(\frac{1}{L^\alpha}\right)$ with $\alpha>1$ as $L\rightarrow \infty$, is a splitting less than that of the
many-body energy levels, which is of order $1/L$, and can indeed be viewed as a
topological degeneracy even for gapless CFT.  The above question can be addressed via
bosonization which maps an interacting fermion gas on a 1$d$ chain to a
free compact boson system on a 1$d$ chain, at low energies.  The
free compact boson system can be solved exactly, and we find that the degeneracy of the ground state is always independent of the interaction. 
More specifically, the Hilbert space of a compact boson with Neumann boundary
condition on an open chain  corresponds to that of a Dirac fermion
with the same left and right boundary condition $\psi_L^i=+\psi^i_R$ $(i=1,2)$ on an open chain.\footnote{More
generally, there are two natural boundary conditions  for a single Majorana fermion: $\psi_L=\eta \psi_R$, $\eta=\pm1$. 
On an open chain, we will choose the same boundary condition on the left end as that of the right end. Therefore, a Dirac
fermion composed of two flavors of Majorana fermions can have four choices of the boundary conditions.
 Here, we choose $\eta=+1$ for both
Majorana fermions and for both boundaries, which corresponds to the Neumann boundary condition after
bosonization.} 
The spectrum does not have nearly degenerate ground state with splitting of
order $O\left(\frac{1}{L^\alpha}\right)$ with $\alpha>1$ as $L\to \infty$, regardless the fermion interaction is repulsive or attractive. We confirm this with the exact computation of the low energy spectrum of the following
spinless interacting fermions on a 1d chain of size $L$,
\begin{align}
 H = \sum_{i=1}^{L-1} (-c_i^\dag c_{i+1}+h.c.)
+ V \left(c_i^\dag c_i -\frac12\right)\left(c_{i+1}^\dag c_{i+1} -\frac12\right)\,.
\end{align} 
After shifting
the ground state energy to zero, we find the spectrum 
$E_n/\left(\frac{\pi}{L+1}\right)$ as follows,
\begin{align*}
\begin{tabular}{ c c | c c c c }
V & L & $n=0$ & $1$ & $2$ & $3$ \\
\hline
-1 & 20  &  0.0~ & 0.439154~ & 0.439154~ & 1.3235 \\
-1 & 10 &  0.0~ & 0.444899~ & 0.444899~ & 1.35815 \\[1ex]
 0 & 20&  0.0~ & 0.999068~ & 0.999068~ & 1.99814 \\
 0 & 10 &  0.0~ & 0.996605~ & 0.996605~ & 1.99321 \\[1ex]
1 & 20 &  0.0~ & 1.65908~ & 1.65908~ & 2.44906 \\
1 & 10 &  0.0~  & 1.61755~ & 1.61755~ & 2.35127 \\
\end{tabular}
\end{align*}
We see that, indeed, regardless
the fermion interaction is repulsive or attractive, the ground state is always
unique, and the energy to reach the first excited state is always of order
$O(\frac1L)$.

The topological transition in the $c=2$ fermionic CFT involves two different
interactions $\cal O$ and ${\cal O}'$.  The transition is forbidden if we
impose the UV symmetry $U(1)_c\times U(1)_{\text{trn}}\times SU(2)$. This
exotic phenomenon may appear in the doped spin-$\fh$ fermionic
models\cite{jiang2018symmetry}.

In this paper, we explored an interesting phenomenon that a $\bZ_2^{\rm IFTO}$ symmetry of a
(1+1)$d$ system is generated by stacking an (1+1)$d$ invertible topological order. Such
a symmetry turns out to  be anomalous as discussed in Section \ref{app:ext}.  

This phenomenon can be quite general.
For bosonic systems, we have invertible topological orders in (2+1)$d$ classified by
$\mathbb{Z}$, and invertible topological orders in (4+1)$d$ classified by
$\mathbb{Z}_2$.  For fermionic systems, we have invertible topological orders in
(0+1)$d$ and (1+1)$d$ classified by $\mathbb{Z}_2$, and invertible topological orders in
(2+1)$d$ classified by $\mathbb{Z}$.  
Therefore, we may have (0+1)$d$ fermionic systems with a $\mathbb{Z}_2$ symmetry generated
by adding a fermion.  Such a symmetry is analogous to  supersymmetry.  In fact, such a (0+1)$d$
fermionic system does exist, which can be formed by two Majorana zero modes at
the two ends of $p$-wave superconducting chain.  Similarly, we may have (1+1)$d$
fermionic systems with a $\mathbb{Z}_2$ symmetry generated by adding a (1+1)$d$
IFTO, which is what we studied in this paper.  We may also have (4+1)$d$ bosonic
systems  with a $\mathbb{Z}_2$ symmetry generated by adding a (4+1)$d$ invertible
bosonic topological order.

\subsection*{Acknowledgments}
We would like to thank N. Benjamin, C. Cordova,  L. Fu, T. Hsieh, P.-S. Hsin,  T. Johnson-Freyd, Z. Komargodski, J. Kulp,  H.T. Lam, M. Metlitski, K. Ohmori, N. Seiberg, T. Senthil, C. Wang, J. Wang, Y. Wang for useful conversations. 
We also thank Y. Tachikawa for comments on a draft. 
The work of  S.H.S.\ is supported  by the National Science Foundation grant PHY-1606531, the Roger Dashen Membership, and  a grant from the Simons Foundation/SFARI (651444, NS). 
X.G.W. is supported by NSF Grant No.\ DMS-1664412. 
W.J.\ thanks the Yukawa Institute for Theoretical Physics at Kyoto University,
where part of this work was completed during the workshop YITP-T-19-03 ``Quantum Information and String Theory 2019.'' 
This work benefited from the 2019 Pollica summer workshop, which was supported in part by the Simons Foundation (Simons Collaboration on the Non-Perturbative Bootstrap) and in part by the INFN. 
S.H.S.\ is grateful for the hospitality of the Physics Department of National Taiwan University  during the completion of this work. 
This research was supported in part by Perimeter Institute for Theoretical Physics. Research at Perimeter Institute is supported by the
Government of Canada through the Department of Innovation, Science and Economic Development and by the Province of Ontario through the
Ministry of Research and Innovation. 
This work was performed in part at Aspen Center for Physics, which is supported by National Science Foundation grant PHY-1607611.

\appendix

\section{Identities for the Arf Invariants}

Here we record some important identities for the Arf invariant and $\bZ_2$ connections (see, for example, \cite{Karch:2019lnn}):
\ie\label{arfid1}
{\rm Arf}[s+t+ \rho] =&{\rm Arf}[s+\rho]+{\rm Arf}[t+\rho]\nn\\
&+{\rm Arf}[\rho]+\int s\cup t\,,
\fe
\ie\label{arfid2}
{1\over 2^g}\sum_{s}
e^{\ii \pi\left({\rm Arf} [s+\rho]  +{\rm Arf} [\rho]+\int s\cup t\right)}= e^{\ii \pi {\rm Arf}[t+\rho]}\,,
\fe
and
\ie\label{complete}
{1\over 2^g} \sum_s e^{\ii \pi \int s\cup t} =  
\begin{cases}
2^g\,,~~~~\text{if}~~t=0\,,\\
0\,,~~~~\text{otherwise}\,.
\end{cases}
\fe

\section{Twisted Torus Partition Functions}

\subsection{Compact Boson}\label{app:ZS1}

In this Appendix we compute the torus partition function of the $S^1$ CFT with non-trivial  $\bZ_2^{\cal B}=\bZ_2^{(1,0)}$ twist  (defined in \eqref{Z2BS1}), and prove   \eqref{KWS1}. 
We will use 0 (1) to denote the trivial (nontrivial) $\bZ_2^{\cal B}$ twist in either the space or time direction. 
In particular, the torus partition function $Z_{S^1}(R)$ \eqref{ZS1} with trivial $\bZ_2^{\cal B}$ background will be denoted as $Z_{S^1}[00]$. 

The torus partition function of $S^1$ with a  $\bZ_2^{\cal B}$ twist in the time direction  is
\ie\label{ZpmS1}
Z_{S^1} [01 ] =  
{1\over |\eta(q)|^2}  \sum_{n,w\in \bZ} (-1)^n q^{ \frac 14 ( {n\over R} +wR)^2   }\, \bar q^{ \frac 14 ( {n\over R} -wR)^2   }\,.
\fe
Next, we  compute the torus partition function $Z_{S^1} [10] $ for the twisted sector of $\bZ_2^{\cal B}$, corresponding to a non-trivial twist along the spatial direction.   
From the discussion in Section \ref{Sec:FreeBoson}, we see that the twisted sector of $\bZ_2^{\cal B}=\bZ_2^{(1,0)}$ are operators $V_{\tilde n,\tilde w}$ with half-integral winding number $w$. Hence
\ie
Z_{S^1} [10 ] =  
{1\over |\eta(q)|^2}  \sum_{\tilde n\in \bZ,\tilde w\in \frac 12+\bZ}  q^{ \frac 14 ( {\tilde n\over R} +\tilde wR)^2   }\, \bar q^{ \frac 14 ( {\tilde n\over R} -\tilde wR)^2   }
\,.
\fe
Finally, the torus partition function with non-trivial $\bZ_2^{\cal B}$ twists in both cycles is the  modular $T$ transform $\tau \to \tau+1$ of  $Z_{S^1} [10 ]$:
\ie
Z_{S^1} [11 ] =  
{1\over |\eta(q)|^2}  \sum_{\tilde n\in \bZ,\tilde w\in \frac 12+\bZ}  (-1)^{\tilde n} q^{ \frac 14 ( {\tilde n\over R} +\tilde wR)^2   }\, \bar q^{ \frac 14 ( {\tilde n\over R} -\tilde wR)^2   }
\,.
\fe
Adding the above four twisted torus partition functions $Z_{S^1}$'s together and dividing by 2, we obtain the torus partition function of the $S^1[R]/\bZ_2^{\cal B}$ theory:
\ie
&Z_{S^1/\bZ_2^{\cal B}}(R)\\
=& \frac12 \Big( \, Z_{S^1}[00](R)+Z_{S^1}[01](R)+Z_{S^1}[10](R)+Z_{S^1}[11](R)\, \Big)\\
=& Z_{S^1}(R/2)\,. 
\fe
Finally by T-duality, $Z_{S^1}(R/2)=Z_{S^1}(2/R)$. Hence, we have  shown \eqref{KWS1} at the level of the torus partition function. 

\subsection{$S^1/\bZ_2$}\label{app:ZS1Z2}

In this Appendix we compute the twisted torus partition functions of $S^1/\bZ_2[R]$ and show \eqref{KWS1Z2}.  
The torus partition function of $S^1/\bZ_2$ with a non-trivial $\bZ_2^{\cal B}$ twist in the time direction  is
\ie\label{Zpm}
Z_{S^1/\bZ_2} [01 ] =  &
{1\over |\eta(q)|^2}  \sum_{n,w\in \bZ} (-1)^n q^{ \frac 14 ( {n\over R} +wR)^2   }\, \bar q^{ \frac 14 ( {n\over R} -wR)^2   }\\
&+\left| {\eta(q) \over \theta_2(q)}\right|\,.
\fe
Note that the twisted sector of $\sigma_1$ contributes oppositely compared to that of $\sigma_2$, so their contribution cancel with each other. 
The other two can be immediately obtained from the modular $S:\tau\to-1/\tau$ and $T:\tau\to\tau+1$ transformations:
\ie
Z_{S^1/\bZ_2} [10 ] = &
{1\over |\eta(q)|^2}  \sum_{n\in \bZ, w\in \frac 12+\bZ}  q^{ \frac 14 ( {n\over R} +wR)^2   }\, \bar q^{ \frac 14 ( {n\over R} -wR)^2   }\\
&+\left| {\eta(q) \over \theta_4(q)}\right|\,\\
Z_{S^1/\bZ_2} [11 ] =  &
{1\over |\eta(q)|^2}  \sum_{n\in \bZ,w\in \frac 12+\bZ}  (-1)^n q^{ \frac 14 ( {n\over R} +wR)^2   }\, \bar q^{ \frac 14 ( {n\over R} -wR)^2   }\\
&+\left| {\eta(q) \over \theta_3(q)}\right|\,.
\fe
Adding the four  $Z_{S^1/\bZ_2}$'s  together and dividing  by 2, we obtain the partition function for ${S^1/\bZ_2 \over \bZ_2^{\cal B}}$:
\ie
&Z_{S^1/\bZ_2 \over \bZ_2^{\cal B}}  (R ) \\
=&
\frac 12 {1\over |\eta(q)|^2} 
\left(  \sum_{n\in2 \bZ, w\in \bZ} + \sum_{n\in  2\bZ,w\in \frac 12+\bZ}  \right)
 q^{ \frac 14 ( {n\over R} +wR)^2   }\, \bar q^{ \frac 14 ( {n\over R} -wR)^2   } \\
&+\left| {\eta(q)\over \theta_2(q)}\right|  +\left| {\eta(q)\over \theta_4(q)}\right|  +\left| {\eta(q)\over \theta_3(q)}\right|  \,.
\fe
Finally, we note that
\ie
&\left(  \sum_{n\in2 \bZ, w\in \bZ} + \sum_{n\in  2\bZ,w\in \frac 12+\bZ}  \right)q^{ \frac 14 ( {n\over R} +wR)^2   }\, \bar q^{ \frac 14 ( {n\over R} -wR)^2   } \\
=&  \sum_{n,w\in \bZ}  q^{ \frac 14 ( {n\over R'} +wR')^2   }\, \bar q^{ \frac 14 ( {n\over R'} -wR')^2   } 
\fe
with $R'={2\over R}$.  Comparing the above with \eqref{Zorb}, we have shown \eqref{KWS1Z2}.

\subsection{$T^2$ CFT}\label{app:ZT2}

The torus partition function of the $T^2$ CFT at a generic point on the conformal manifold is given by
\ie\label{T2pp}
Z_{\cal B} [00] = \text{Tr}_{\cal H} [q^{L_0 -{c\over24}  }  \bar q^{ \bar L_0 -{c\over24} }] = 
 {1\over |\eta(q)|^4}\sum_{n_i,w^i\in \bZ}  q^{h -{1\over12}}\bar q^{\bar h-{1\over12}} 
\fe
where $h,\bar h$ are given as in \eqref{T2hh}.

The torus partition function with a $\bZ_2^{\cal B}$ twist along the time direction is
\ie\label{T2pm}
Z_{\cal B}[01] =& \text{Tr}_{\cal H} [ \eta \, q^{L_0 -{c\over24}  }  \bar q^{ \bar L_0 -{c\over24} }]\nn\\
=&  {1\over |\eta(q)|^4}\sum_{n_i,w^i\in \bZ}  (-1)^{n_1+w^1+n_2+w^2} \,q^{h -{1\over12}}\bar q^{\bar h-{1\over12}} \,.
\fe

Next, we consider the torus partition function with a $\bZ_2^{\cal B}$ twist in the spatial direction, which counts non-local operators living in the twisted sector $\widetilde{\cal H}$.   These non-local operators are of the form \eqref{O} but with fractional momentum and winding numbers $\tilde n_i,\tilde w^i$:
\ie
V _{\tilde n_1,\tilde w^1,\tilde n_2,\tilde w^2}:~~\tilde n_i,\tilde w^i \in \frac 12 +\bZ\,.
\fe
Note that the spin $s=\tilde n_i \tilde w^i$ of such operator is either an integer or a half integer, consistent with the spin selection rule for a non-anomalous $\bZ_2^{\cal B}$ \cite{Hung:2013cda,Chang:2018iay,Lin:2019kpn}.  
The $\bZ_2^{\cal B}$ charge of $V_{\tilde n_1,\tilde w^1,\tilde n_2,\tilde w^2}$ is given by $(-1)^{2s}$. 
The partition function with a $\bZ_2^{\cal B}$ twist in the time direction is therefore:
\ie\label{T2mp}
Z_{\cal B}[10]   =& \text{Tr}_{\widetilde{\cal H}} [  \, q^{L_0 -{c\over24}  }  \bar q^{ \bar L_0 -{c\over24} }]\nn\\
= & {1\over |\eta(q)|^4}\sum_{\tilde n_i,\tilde w^i\in\frac12+ \bZ}   \,q^{h -{1\over12}}\bar q^{\bar h-{1\over12}}  \,.
\fe
Finally, the torus partition function with $\bZ_2^{\cal B}$ twists  both in the spatial and time direction is
\ie\label{T2mm}
Z_{\cal B}[11]  =& \text{Tr}_{\widetilde{\cal H}} [ \eta \, q^{L_0 -{c\over24}  }  \bar q^{ \bar L_0 -{c\over24} }]\nn\\
=&  {1\over |\eta(q)|^4}\sum_{\tilde  n_i,\tilde w^i\in\frac12+ \bZ}   \, (-1)^{2\tilde n_i \tilde  w^i}\,q^{h -{1\over12}}\bar q^{\bar h-{1\over12}}  \,.
\fe

At the special point of ${\mathfrak{su}(2)}_1\times {\mathfrak{su}(2)}_1$, the torus partition functions are 
\ie
&Z_{\cal B}[00]  = \left(\,  | \chi_{0}^{su2_1}|^2+| \chi_{1\over2}^{su2_1}|^2 \,\right)^2\,,\\
&Z_{\cal B}[01]  = \left(\,  | \chi_{0}^{su2_1}|^2  -  | \chi_{1\over2}^{su2_1}|^2 \,\right)^2\,,\\
&Z_{\cal B}[10]  = \left(\,   \chi_{0}^{su2_1} \bar \chi_{1\over2} ^{su2_1} + \chi_{1\over2}^{su2_1} \bar \chi_{0} ^{su2_1} \,\right)^2\,,\\
&Z_{\cal B}[11]  = - \left(\,   \chi_{0}^{su2_1} \bar \chi_{1\over2} ^{su2_1} - \chi_{1\over2}^{su2_1} \bar \chi_{0} ^{su2_1} \,\right)^2\,,
\fe
where $\chi^{su2_1}_j(q)$  is the $\mathfrak{su}(2)_1$ current algebra characters with $\mathfrak{su}(2)$ spin $j$. 
At level 1, there are only two allowed spins, $j=0$ and $j=1/2$, whose conformal weights $h$ are 0 and $1/4$, respectively. Their characters are $\chi^{su2_1}_0 (q)  =  {\theta_3(q^2 ) \over \eta(q)}\,,~\chi^{su2_1}_{1\over2} (q )  =  {\theta_2(q^2) \over \eta(q)}$. 
The torus partition function of the fermionized theory, {\it i.e.} four Majorana fermions, with respect to the four spin structures are
\ie
&Z_{\cal F} [AA] = 
\left[\, (\chi^{su2_1}_{0})^2  + ( \chi^{su2_1} _{1\over2} )^2 \,\right]
\left[\, (\bar\chi^{su2_1}_{0})^2  + (\bar \chi^{su2_1} _{1\over2} )^2 \,\right] \,,\\
&Z_{\cal F} [AP] = 
\left[\, (\chi^{su2_1}_{0})^2  -( \chi^{su2_1} _{1\over2} )^2 \,\right]
\left[\, (\bar\chi^{su2_1}_{0})^2   -(\bar \chi^{su2_1} _{1\over2} )^2 \,\right] \,,\\
&Z_{\cal F} [PA] = 4 \, \chi^{su2_1}_{0} \chi^{su2_1} _{1\over2}  \bar\chi^{su2_1}_{0} \bar\chi^{su2_1}_{1\over2 }\,,\\
&Z_{\cal F}[PP]=0\,.
\fe

\section{The Tomonaga-Luttinger Liquid Theory}\label{app:TL}
In this appendix we distinguish two different duality maps in the Tomonago-Luttinger (TL) model\cite{Tomonaga:1950zz,Luttinger:1963zz,Haldane:1981zza}. The TL liquid is described by the following action, in the imaginary time with coordinate $(x^1, x^2)=(x,-\ii t)$, 
\begin{align}
S_{\text{TL}}=&\frac{v_F}{2\pi}\int d^2x \left[ \frac{1}{K}(\partial_1 \phi)^2+K(\partial_1 \theta)^2\right] \nn\\
&-\frac{\ii}{\pi}\int d^2x\partial_1 \theta \partial_2 \phi
\end{align}
The canonical commutation relation is
\begin{align}
\left[\partial_1 \phi (x), \frac{1}{\pi}\partial_2 \theta (x')\right] =\ii \delta (x-x')\label{dualfield}
\end{align}
which is independent of $K$. It follows that when $K=1$, the theory is equivalent to the bosonized theory of the free Dirac fermion.

From this action, one can integrate out $\theta$ to obtain an action for the $\phi$ field,
\begin{align}
S_{\text{TL}}[\phi]=\frac{1}{2\pi K v_F}\int d^2x \left[ (\partial_2 \phi)^2+v_F^2 (\partial_1 \phi)^2\right] 
\end{align}
where 
\begin{align}
\phi (x^1+2\pi, x^2)= \phi (x^1)+2\pi R_\phi Q_\phi,\quad Q\in\ZZ
\end{align}
with $R_\phi=\sqrt{2}K$, normalized such that the free Dirac theory with $K=1$ has $R_\phi=\sqrt{2}$. This is nothing but the action of free boson CFT $S^1[R]$ with compactification radius $R=R_\phi$. The Luttinger liquid theory of the interacting spinless fermion is, more precisely, the fermionized theory ${\rm Dirac}[R]$. 

Alternatively, one can integrate out $\phi$ to obtain an action for the $\theta$ field,
\begin{align}
S_{\text{TL}}[\theta]=\frac{K}{2\pi  v_F}\int d^2x \left[ (\partial_2 \theta)^2+v_F^2 (\partial_1 \theta)^2\right] 
\end{align}
where 
\begin{align}
\theta (x^1+2\pi, x^2)= \theta (x^1)+2\pi R_\theta Q_\theta,\quad Q_\theta \in\ZZ
\end{align}
with $R_\theta=\frac{\sqrt{2}}{K}$. The normalization is to be consistent with (\ref{dualfield}). The field $\phi$ and dual field $\theta$ can be equivalently represented by chiral fields $\phi (z,\bar z)= X_L(z)+X_R(\bar z)$ and $\theta (z,\bar z)=X_L (z)-X_R(\bar z)$. 

There are two distinct maps along the moduli space of $c=1$ CFT parametrized by the radius $R=R_\phi$. 
\begin{itemize}
\item T-duality:
\begin{align}
R\rightarrow \frac{1}{R}, ~~ X_L\rightarrow X_L, ~~ X_R\rightarrow -X_R\,,
\end{align}
under which the bosonic operator $V_{n,w}(R)\rightarrow V_{w,n} \left( \frac{1}{R}\right)$. 
Namely, the T-duality exchanges the momentum $n$ and the winding $w$ numbers of local operators $V_{n,w}$ in the bosonic CFT. 
In terms of  the fields in the Luttinger model, the T-duality  maps  $\phi\rightarrow\theta, R_{\phi}\rightarrow \frac{1}{R_\phi}\neq R_\theta $. Under T-duality, the free Dirac point with $R_\phi=\sqrt{2}$ is \emph{not} the self-dual point. Instead, the $\mathfrak{su}(2)_1$ CFT with $R=1$ is the self-dual point. 

\item IFTO-stacking:
\begin{align}
R\rightarrow \frac{2}{R}, ~~ X_L\rightarrow X_L, ~~ X_R\rightarrow -X_R\,,
\end{align}
under which the fermionic operator $V_{n,w}(R)\rightarrow V_{-w,-\frac{n}{2}} \left( \frac{2}{R}\right)$. 
The corresponding map in the Luttinger model is $K\rightarrow \frac{1}{K},\phi\rightarrow \theta, R_\phi\rightarrow R_\theta$. 
Under this map, the Dirac theory \emph{is} the self-dual point \cite{Karch:2019lnn}.
\end{itemize}

\bibliography{EMPT,all,publst}

\end{document}